\documentclass[useAMS, usenatbib]{mn2e}
\usepackage{aas_macros}
\usepackage{amsmath}
\usepackage{amssymb} 
\usepackage{graphicx} 
\usepackage{float} 
\usepackage{amsfonts}
\usepackage{lmodern}
\usepackage{bm}
\usepackage{placeins}
\usepackage{color}
\usepackage{booktabs} 
\usepackage{multirow}
\usepackage{gensymb}
\usepackage{hyperref}
\usepackage{enumerate}
\usepackage[export]{adjustbox}

\newcommand{\bc}{\begin{center}}
\newcommand{\ec}{\end{center}}

\newcommand{\GG}[1]{}

\let\oldhat\hat
\renewcommand{\hat}[1]{\oldhat{\mathbf{#1}}}
\let\oldbullet\bullet \renewcommand{\bullet}[1][0pt]{%
\mathrel{\raisebox{#1}{$\oldbullet$}}%
}
\title[AGN-driven outflows in simulated dwarfs]{Fast and energetic AGN-driven outflows in simulated dwarf galaxies} \author[Koudmani, Sijacki, Bourne \& Smith]{Sophie Koudmani$^1$\thanks{E-mail: skoudmani@ast.cam.ac.uk}, Debora
  Sijacki$^1$, Martin A. Bourne$^1$ and Matthew C. Smith$^{1,2}$ \\
  $^1$ Institute of Astronomy and Kavli Institute for Cosmology,
  University of Cambridge, Madingley Road, Cambridge CB3 0HA, UK \\
  $^2$ Center for Computational Astrophysics, Flatiron Institute, 162 5th Avenue, New York, NY 10010, USA}

\begin{document}

\maketitle

\begin{abstract}
The systematic analysis of optical large-scale surveys has revealed a population of dwarf galaxies hosting AGN, which have been confirmed by X-ray follow-up observations. Recently, the MaNGA survey identified six dwarf galaxies that appear to have an AGN that is preventing on-going star formation. It is therefore timely to study the physical properties of dwarf galaxies, in particular whether the presence of an AGN can affect their evolution. Using the moving mesh code \textsc{arepo}, we have investigated different models of AGN activity, ranging from simple energy-driven spherical winds to collimated, mass-loaded, bipolar outflows in high resolution simulations of isolated dwarf galaxies hosting an active black hole. Our simulations also include a novel implementation of star formation and mechanical supernova (SN) feedback. We find that AGN outflows have a small but systematic effect on the central star formation rates (SFRs) for all set-ups explored, while substantial effects on the global SFR are only obtained with strong SNe and a sustained high-luminosity AGN with an isotropic wind. This suggests that AGN feedback in dwarf galaxies is unlikely to directly regulate their global SFRs. There is, however, a significant effect on outflow properties, which are notably enhanced by the AGN to much higher outflow temperatures and velocities, in agreement with kinematic signatures from the MaNGA survey. This indicates that AGN may play an indirect role in regulating the baryon cycle in dwarf galaxies by hindering cosmic gas inflows.
\end{abstract}

\begin{keywords}
methods: numerical, galaxies: formation, galaxies: dwarf, galaxies: active
\end{keywords}

\section{Introduction} \label{intro}
In the $\Lambda$CDM Universe, structure formation is a hierarchical process whereby primordial density fluctuations collapse into dark matter haloes with cosmic time. The gravitational pull of the dark matter leads the gas to collapse within these haloes, where it can radiatively cool and form stars. Dwarf galaxies, with stellar masses of $M_{\ast} \lesssim 3 \times 10^9\ \mathrm{M_{\odot}}$ (or approximately the mass of the Large Magellanic Cloud), are the smallest building blocks of this structure formation process. They are interesting cosmological probes, as there are a number of apparent discrepancies when comparing observations of dwarf galaxies with the predictions of the $\Lambda$CDM model.

These discrepancies concern, for example, a large disparity in the number of observed dwarfs with respect to the predicted number of dark matter haloes that may be hosting these systems - the so-called missing satellite problem \citep[e.g.][]{Kauffmann1993, Klypin1999,Moore1999}. There is also a lack of observed massive satellites - the too-big-to-fail problem \citep{Boylan-Kolchin2011}. Furthermore, the inferred dark matter halo profiles from observations of dwarf galaxies do not seem to agree with the cuspy dark matter profile predicted by the $\Lambda$CDM model - the cusp vs. core problem \citep[for one of the first papers on this subject, see][]{Moore1994}.

This has prompted some to question the validity of $\Lambda$CDM and seek alternatives such as warm dark matter \citep[e.g.][]{Lovell2012} or self-interacting dark matter \citep[e.g.][]{Vogelsberger2014a}. On the other hand, a large body of theoretical work has identified a number of important baryonic processes which may fundamentally affect dwarf properties and largely alleviate the aforementioned discrepancies \citep[see e.g.][]{Pontzen2014}.

The first process concerns photoheating from the metagalactic UV background which drastically reduces the number of luminous low-mass haloes, with a characteristic halo mass for reionization suppression of $M_{c}  \sim 10^{10}\ \mathrm{M_{\odot}}$ at $z=0$ \citep[e.g.][]{Efstathiou1992,Okamoto2008,Fitts2017}. The second process regards feedback from SNe, which is commonly invoked in simulations as it is one of the energetically most promising mechanisms \citep[e.g.][]{Dekel1986}. In the APOSTLE simulations \citep{Sawala2016} reionization and SN feedback allow galaxy formation to proceed only in a small subset of dark matter haloes.  Other groups find that additional physical processes (e.g. photoionization, radiation pressure, or stellar winds) are required for SNe to couple to the interstellar medium (ISM) effectively and drive outflows \citep[see e.g.][]{Garrison-Kimmel2014a,Hopkins2014,Kimm2015,Smith2018b, Emerick2018}. It is also yet unclear whether SN feedback could induce a change in the dark matter distribution from cusp to core \citep[see e.g.][]{Navarro1996}. For example, \citet{Governato2010} and \citet{Parry2012} observe this effect in their simulations, while others only find cuspy profiles \citep{Vogelsberger2014a,Sawala2016}.

The importance of other feedback channels in dwarf galaxies is still controversial. Recently, it has been suggested by \citet{Silk2017} that AGN feedback from intermediate mass black holes (IMBHs) in dwarf galaxies could solve many of the above issues including abundances, baryon fraction, and the cusp vs. core problem. For massive galaxies, AGN feedback has been included as a canonical component of the galaxy formation model \citep{Binney1995,DiMatteo2005,Bower2006,Croton2006, Sijacki2007}, since virtually all massive galaxies are known to host a supermassive black hole with $10^6\ \mathrm{M_{\odot}}<M_\mathrm{BH}<10^{10}M_{\odot}$ at their centre \citep[for a recent review see][]{Kormendy2013}. AGN feedback can bring the stellar properties in agreement with observations for the high-mass end of the galaxy population \citep{Puchwein2013,Vogelsberger2014,Schaye2015,Beckmann2017,Henden2018} and various simulations have explored different ways of how AGN feedback could couple to the gas \citep[see e.g.][]{DeBuhr2010,Nayakshin2012,Choi2012,Choi2015,Costa2014,Costa2018}. In the past few years, there has been growing observational evidence that AGN feedback could also be important at the low-mass end of the galaxy population. 

Some of the first systematic searches for AGN in dwarf galaxies were carried out by searching the Sloan Digital Sky Survey \citep[SDSS;][]{York2000} for broad H$\alpha$ emission lines \citep{Greene2004, Greene2007}. Further X-ray studies confirmed the AGN nature of these sources \citep{ Desroches2009,Dong2012}. However, note that their sample had a median black hole mass of $M_\mathrm{BH}=1.3\times10^6\ \mathrm{M_{\odot}}$ which is still significantly above the IMBH mass regime ($M_\mathrm{IMBH} \leq 10^{5}\ \mathrm{M_{\odot}}$). \citet{Reines2013} managed to extend this sample to lower black hole masses by searching for narrow-line AGN signatures in SDSS identifying 136 candidates with masses in the range $10^5\ \mathrm{M_{\odot}}<M_\mathrm{BH}<10^{6}\ \mathrm{M_{\odot}}$. Follow-up studies in the optical studies found velocity dispersions $\sigma_{\ast}= 20-71\ \mathrm{kms^{-1}}$ and Eddington fractions, $f_\mathrm{Edd}$, from $0.1-50$ per cent, and (where detected) nuclear X-ray luminosities were significantly higher than expected from X-ray binaries \citep{Baldassare2016, Baldassare2016a}. 

\citet{Pardo2016} identified 605 dwarf galaxies at redshift $z<1$ in the NEWFIRM Medium Band Survey \citep{Whitaker2011} that lie within the region covered by archival \textit{Chandra} data. They find an AGN occupation fraction of $0.6-3$ per cent in the stellar mass range $10^9\ \mathrm{M_{\odot}}\leq M_{\ast} \leq 3 \times 10^{9}\ \mathrm{M_{\odot}}$ consistent with local studies, e.g. \citet{Lemons2015}. There have also been a number of systematic searches for IMBHs in the infrared band, which yielded similar AGN occupation fractions  \citep{Satyapal2007, Satyapal2008, Satyapal2014,Marleau2013,Sartori2015}. 

\citet{Chilingarian2018a} carried out an automated search analysing one million SDSS galaxy spectra. They only keep objects with $M_\mathrm{BH} < 2 \times 10^{5}\ \mathrm{M_{\odot}}$ and the BPT \citep{Baldwin1981} classification has to be ``AGN" or ``composite", among other criteria. This yields a sample of 305 IMBH candidates including all previously known AGN in the IMBH mass regime. 18 of those have sufficiently deep X-ray observations to rule out or confirm an AGN. This results in a sample of 10 `bona fide' intermediate mass AGN, five of which were previously known. Furthermore, IMBHs with masses down to $\sim10^{3}\ \mathrm{M_{\odot}}$ have been predicted based on known scaling relations for both early-type \citep{Graham2018b} and late-type \citep{Graham2018a} galaxies in the Virgo Cluster.

More direct evidence for AGN feedback comes from the SDSS-IV \citep{Blanton2017} MaNGA \citep{Bundy2015} survey where 6 out of 69 quenched low-mass ($M_{\ast}<5\times10^9\ \mathrm{M_{\odot}}$) galaxies appear to have an active AGN that is preventing on-going star formation \citep{Penny2018a}. Five of these six galaxies have kinematic signatures consistent with outflows. This discovery challenges the canonical picture of quenching mechanisms where it is assumed that stellar feedback channels are solely responsible for regulating star formation in low-mass galaxies. Might AGN also play a role in the evolution of dwarf galaxies?

Dashyan et al. (2017) address this question using an analytical model and find that AGN feedback is more efficient than SN feedback for driving gas out of dwarf galaxies for a significant range of parameter space. However, their model includes some simplifying assumptions (e.g. spherical symmetry) and to settle this issue a more comprehensive numerical model is needed. \citet{Wadepuhl2010} carried out high-resolution hydrodynamic simulations of the formation of Milky Way sized galaxies and their satellites to test whether including additional feedback mechanisms could address the `missing satellite problem'. They find that including AGN feedback in addition to SN feedback does not make a significant difference as most of the satellites are too small to grow a large black hole.

Mini-quasars have been invoked as direct contributors to reionization for many years. Though now it seems that massive stars play the most important role in the reionization process producing the bulk of UV photons, while the contribution to the ionization fraction from X-rays is only a few per cent \citep{Ciardi2010}. \citet{Trebitsch2017} carried out cosmological zoom-in simulations of high-redshift dwarf galaxies to determine whether an AGN could have an indirect effect on reionization by creating paths for photons to escape. However, they find that SN feedback is far more efficient than AGN feedback in their simulations as the AGN is starved of gas resulting in low accretion rates. Similarly, \citet{Habouzit2017} find that in their cosmological simulations of high-redshift dwarf galaxies, the growth of BHs is hindered by strong SN feedback. However, other high-redshift simulations find that Eddington fractions increase with time to $0.2-0.8$ at redshift $z=4$ \citep{Barai2018a}. \citet{Bellovary2018} set up cosmological zoom-in simulations focussing on dwarf galaxies at $z=0$. They also find that accretion rates are low throughout cosmic time rendering them difficult to detect electromagnetically. But they note that their merger history is optimal for gravitational wave detection by LISA.

For most of the simulations mentioned above, the Bondi-Hoyle-Lyttleton (BHL) accretion rate \citep{Hoyle1939,Bondi1944} is used as the AGN accretion prescription which, due to its quadratic dependency on the black hole mass, makes it difficult for IMBHs to accrete significant amounts of gas.  As feedback is directly coupled to accretion, this will then also result in low AGN luminosities. Note that whilst this approach allows simulators to model accretion self-consistently, there is no observational evidence for central black holes accreting at the BHL rate. Moreover, observations of intermediate mass AGN consistently find high Eddington fractions so there must be a scenario which allows IMBHs to accrete efficiently.

Here we therefore decide to take a phenomenological, but more adaptable approach and let the AGN in our simulations accrete at a fixed fraction of the Eddington rate $\dot{M}_\mathrm{Edd}$ based on the initial black hole mass. This allows us to explore a wide parameter space to determine which AGN luminosity would be required for AGN feedback to affect dwarf galaxies and then compare back to observations to assess whether such luminosities would be realistic\footnote{Note that \citet{Trebitsch2017} also have one run in their simulation suite where they let the BH accrete as much gas `as possible' limited by the Eddington rate. However, in practice their accretion algorithm does not allow for significant accretion once the central region gets depleted by SN feedback and therefore the accretion ends up being sub-Eddington for the majority of the simulation.}.
The aim is to assess the maximum possible impact of AGN feedback, without worrying how the gas would get accreted, and investigate the effect on the baryonic content of dwarf galaxies, including the gas content, star formation and outflow properties.

The outline of the paper is as follows. We describe our numerical methods in Section \ref{section:Methodology}, including a summary of the SN and AGN models. In Section \ref{section:Sims}, we describe the different set-ups for our isolated dwarf galaxy simulations, in particular the initial conditions and the `switched-on' physics modules for each simulation run. We then present our simulation results in Section \ref{section:Results} where we analyse the star formation and outflow properties, and compare our results to observations from MaNGA. We discuss these results and compare them to other studies in Section \ref{section:Discussion}, and finally conclude in Section \ref{section:Conclusion}.
\section{Methodology} \label{section:Methodology}
All simulations presented in this work are carried out with the massively-parallel \textsc{arepo} code \citep{Springel2010}, where fluids are modelled as a moving mesh using a quasi-Lagrangian finite volume technique. The unstructured mesh is based on the Voronoi tessellation of a set of discrete points that cover the whole computational domain. These mesh-generating points are allowed to move with the local flow velocity, with minor corrections to avoid excessive distortion of the gas cells. 

Gravitational interactions are modelled using the TreePM approach, with star particles as collisionless particles representing whole stellar populations.  For our isolated set-ups we simply model dark matter as a static gravitational potential. The sub-grid models for SN and AGN feedback physics are extensively described in \citet{Smith2018} and \citet{Curtis2015} and only a brief summary is given below (see Sections \ref{SFmodel} and \ref{AGNmodel}). In addition to these, we include primordial and metal line cooling \citep[cf.][]{Vogelsberger2013} and we do not include a UV background. Note that we impose a non-thermal pressure floor to ensure that the Jeans length is resolved by a minimum number of cells, $\mathrm{N_{J,min}= 14}$, to avoid artificial fragmentation \citep[see][for details]{Smith2018}.
\subsection{Star Formation \& ISM Model} \label{SFmodel}
\subsubsection{Star Formation}
If a gas cell's density is above a certain threshold $n_\mathrm{SF}$, it is marked as star-forming and its SFR is calculated using a simple Schmidt law:
\begin{equation}\label{eq:SchmidtLaw}
\dot{\rho}_{\ast} = \epsilon_\mathrm{SF} \frac{\rho}{t_\mathrm{ff}},
\end{equation}
where $ \epsilon_\mathrm{SF}$ is the instantaneous star formation efficiency, $\rho$ is the gas density and $t_\mathrm{ff} = \sqrt{3\pi / 32G\rho}$ is the dynamical free-fall time. For the density threshold we take a fiducial value of  $n_\mathrm{SF}=10\ \mathrm{cm^{-3}}$. There is indication of a wide range of star formation efficiencies in molecular clouds ranging from less than 1 to 40 per cent \citep[see e.g.][]{Krumholz2007}, so we test two different values for the instantaneous star formation efficiency: $ \epsilon_\mathrm{SF} = 1.5$ per cent and $ \epsilon_\mathrm{SF} = 15$ per cent (see Section \ref{feed_config} for a more detailed discussion of the choice of the $ \epsilon_\mathrm{SF}$ parameter).
\subsubsection{SN Feedback}
With this feedback scheme, individual SNe are resolved in time and at each time-step for each star particle SN rates, $\dot{N}_\mathrm{SN}$, are tabulated as a function of age and metallicity from \textsc{Starburst99} \citep{Leitherer1999} assuming a \citet{Kroupa2002} initial mass function. The number of SNe of a given particle in a given time-step then gets drawn from a Poisson distribution with mean $\bar{N}_\mathrm{SN}=\dot{N}_\mathrm{SN} \Delta t $, where $\Delta t$ is the time-step. Note that we also introduce a time step limiter so that $\bar{N}_\mathrm{SN} \ll 1$.

For each SN, mass, metals, energy and momentum are distributed into the gas cell hosting the star particle as well as all neighbouring gas cells that share a face with the host cell. To achieve an isotropic distribution of these quantities, we use the weighting scheme from \citet{Hopkins2018, Hopkins2018a} assigning a vector weight $\mathbf{\bar{w}_{i}}$ to each neighbour \citep[see also][]{Smith2018}. The host cell receives a set fraction $f_\mathrm{host}=0.05$ of the ejecta mass $m_\mathrm{ej}$ and the SN energy $E_\mathrm{SN}$. Here we adopt $m_\mathrm{ej}=10\ \mathrm{M_{\odot}}$, $2 \ \mathrm{M_{\odot}}$ in metals, and $E_\mathrm{SN}=10^{51}$ erg. The total momentum imparted to the neighbouring cells is $p_\mathrm{tot}=\sqrt{2 m_\mathrm{ej} E_\mathrm{SN}}$.

In the `no feedback' simulations, host cells and neighbours receive mass and metals as described above but their energy and momentum are not altered. For other simulations, we use the `mechanical feedback' mode which aims to account for the $P\mathrm{d}V$ work done during the Sedov-Taylor phase of the SN remnant (SNR) evolution in the event that it is unresolved. During this adiabatic phase, the momentum of the SNR can be boosted by around an order of magnitude. By comparing the initial momentum of the SNR to its momentum after it has swept up some ambient material, this boost factor can be parametrized in terms of the weighted ejecta mass $\Delta m_{i}=|\mathbf{\bar{w}_{i}}| m_\mathrm{ej}$ and the swept-up cell mass $m_{i}$:
\begin{equation}\label{eq:boostfactor}
f_\mathrm{boost} = \mathrm{MIN}\left[ \sqrt{1+\frac{m_{i}}{\Delta m_{i}}}, \frac{p_\mathrm{fin}}{p_\mathrm{tot}}\right], 
\end{equation}
where $p_\mathrm{fin}$ is the momentum of the SNR as it exits the energy conserving Sedov-Taylor phase and transitions to the momentum conserving snowplough phase. From fits to simulations of individual SNe \citep[see][]{Blondin1998, Thornton1998, Geen2015, Kim2015, Martizzi2015, Kimm2015}, we take the asymptotic momentum as:
\begin{equation}\label{eq:finmomentum}
p_\mathrm{fin} = 3 \times 10^5 \mathrm{km\ s^{-1}\ \mathrm{M_{\odot}}} E_{51}^{16/17} n_\mathrm{H}^{-2/17} Z'^{-0.14},
\end{equation}
where $E_{51}=(E_\mathrm{SN}/10^{51}\mathrm{erg})=N_\mathrm{SN}$ is the number of SNe, $Z'=\mathrm{MAX}(Z/\mathrm{Z_{\odot}}, 0.01)$ is the metallicity in solar units, and $n_\mathrm{H}$ is the hydrogen number density in $\mathrm{cm^{-3}}$. Note that this boost factor is calculated for each neighbouring cell individually.
\subsection{AGN model} \label{AGNmodel}
\subsubsection{Refinement strategy}
\begin{figure*}
\centering
\includegraphics[width=0.4\textwidth]{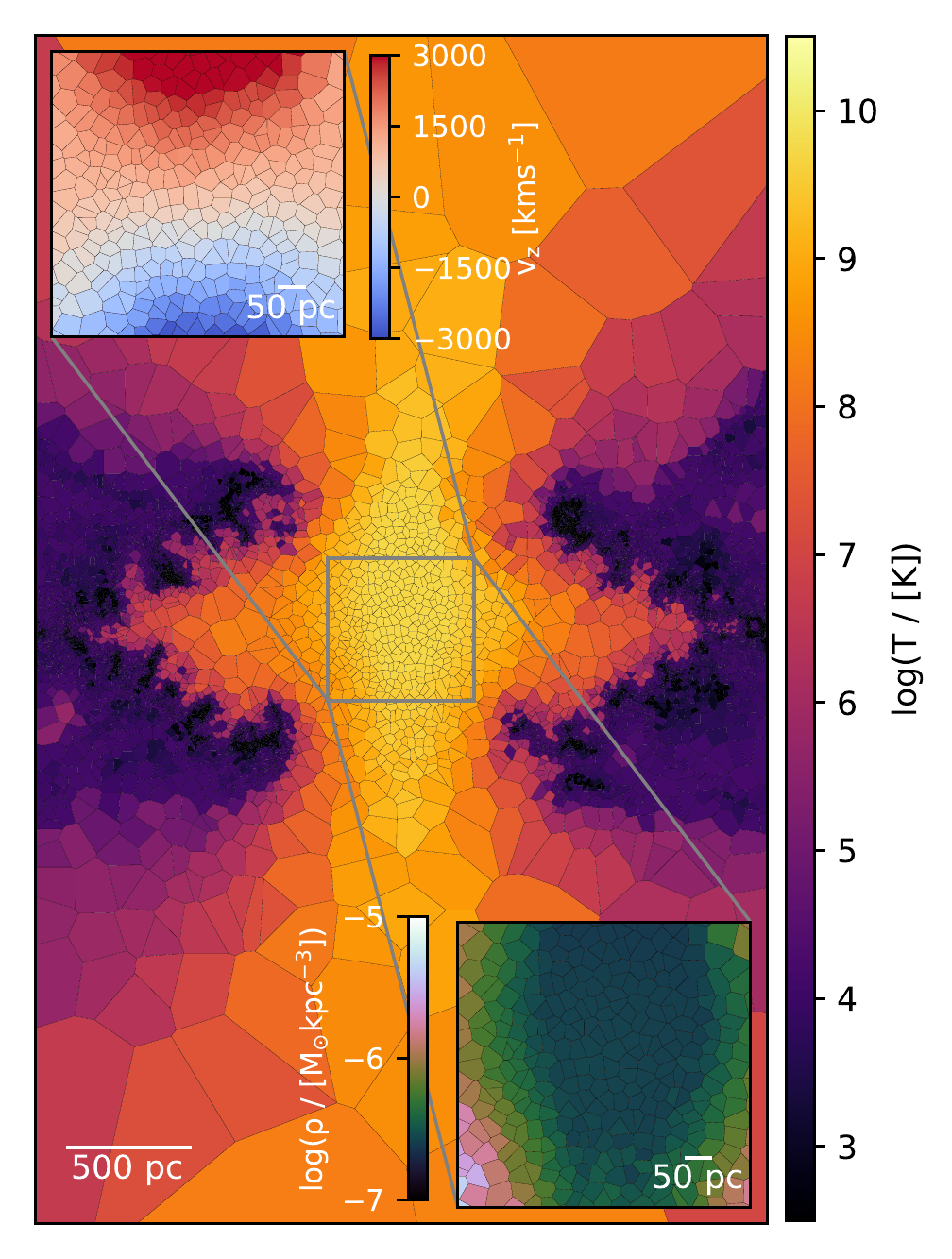}
\includegraphics[width=0.4\textwidth]{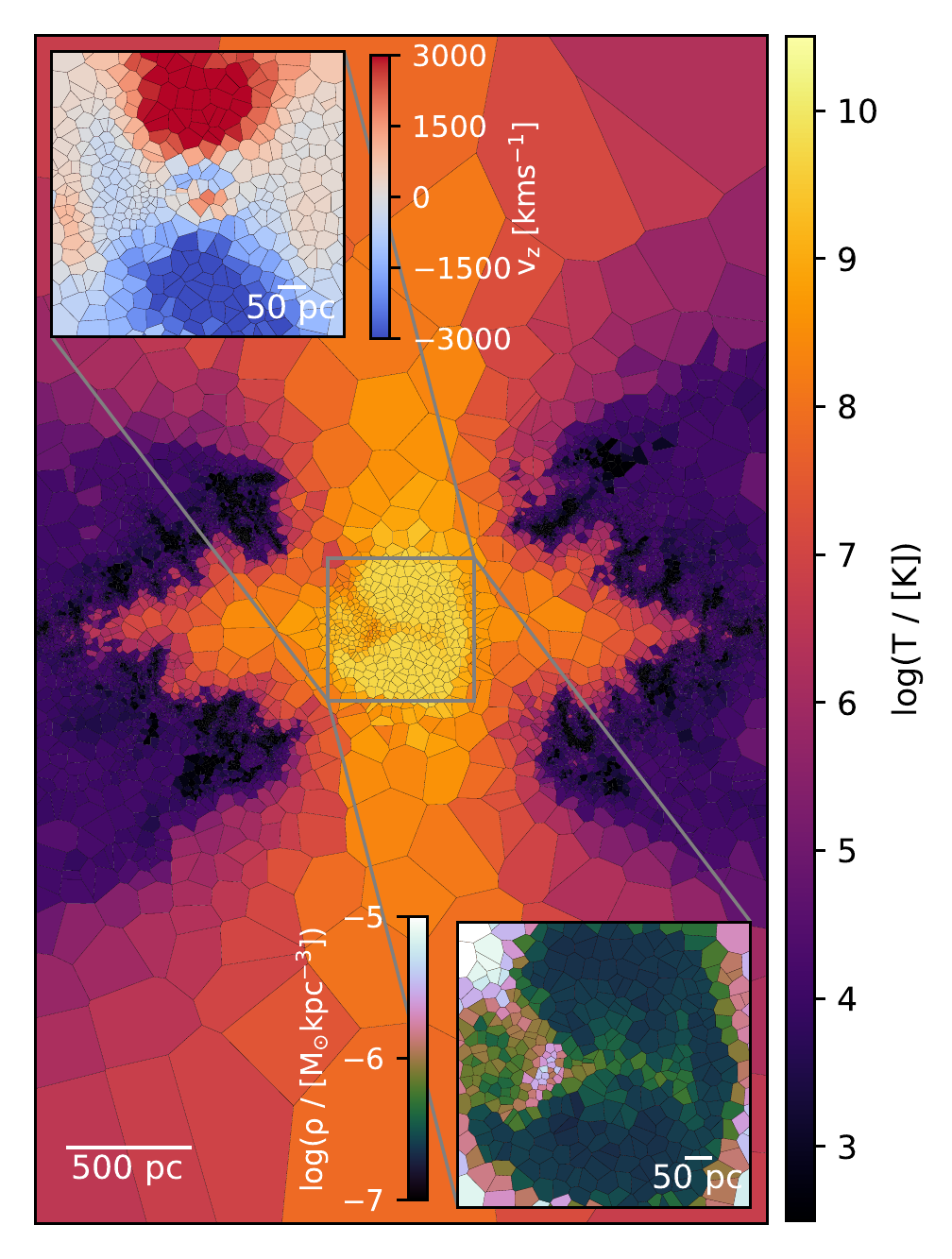}
\caption{Edge-on projections of the Voronoi mesh ($3.0\times5.0\times0.2$ kpc) for the two different AGN injection models at $t\sim25$ Myr. The left panel shows the isotropic, thermal feedback scheme, and the right panel shows the bipolar, mass-loaded outflows. The colour coding in the main panels indicates the gas temperature, and the colour coding in the two insets indicates the vertical gas velocity and gas density, respectively.}
\label{feed_injection} 
\end{figure*}
\textsc{arepo} is a quasi-Lagrangian code, i.e. it aims to keep cell masses constant within a factor two of the target mass $m_\mathrm{target}$. However, alternative refinement criteria can be implemented allowing for regions of interest to be refined to much lower masses and higher spatial resolution. The AGN model by \citet{Curtis2015} uses a super-Lagrangian refinement criterion for the region around the black hole with the aim of resolving gas flows, and therefore AGN feedback mechanisms, more accurately.  The size of the refinement region is set equal to the black hole smoothing length $h_\mathrm{BH}$ which is defined as the radius of the sphere containing a total mass of $32\times m_\mathrm{target}$. This ensures that the refinement region contains a constant amount of mass. Cells within the refinement region then get (de-)refined so that their cell radius $R_\mathrm{cell}$ is within 
\begin{gather}
	0.5 l_\mathrm{ref} < R_\mathrm{cell} <  l_\mathrm{ref},\\
	l_\mathrm{ref} = \frac{d}{R_\mathrm{cell}} (R_\mathrm{cell,max}-R_\mathrm{cell,min}),
\end{gather}
where $d$ is the gas cell's distance from the black hole, $R_\mathrm{cell,min}$ is the minimum cell radius (set to the Bondi radius $r_\mathrm{B}$), and $R_\mathrm{cell,max}$ is the maximum cell radius (set to $0.1h_\mathrm{BH}$). Note that in the original implementation $R_\mathrm{cell,max}$ is set to $0.5h_\mathrm{BH}$ though we find that in our simulations we need a lower value to obtain a relatively smooth cell size distribution (see Appendix \ref{SuperLagRef}). We also set a minimum gas cell mass $m_\mathrm{min}=2\times10^{-1}\ \mathrm{M_{\odot}}=10^{-3}m_\mathrm{target}$. Note that this means that for the majority of the simulation, the Bondi radius is not resolved. However, this is not a significant concern as we have constant accretion rates, and therefore only need outflow properties to be converged. We test both higher and lower values for the minimum mass and find that for resolving energy and mass injection, the above minimum mass value is sufficient (see Appendix \ref{min_mass}). 

For feedback injection, we use both an isotropic scheme with simple energy-driven spherical winds and a non-isotropic scheme with collimated, mass-loaded, bipolar outflows.  The latter is motivated by unresolved dynamics, e.g. magnetic fields close to the black hole, which lead to the observed bipolar outflows from AGN \citep{Rupke2011, Maiolino2012}. Note that we fix the bipolar cone's orientation to be aligned with the z axis. 

The two AGN models are depicted in Figure \ref{feed_injection}. The main panels show edge-on projections of the Voronoi mesh with the colour coding indicating the temperature of the gas cells. Each panel also has two insets zooming in on the central region with colour coding representing the gas velocity and the gas density, respectively. We notice much stronger and more collimated outflows for the bipolar injection model with high gas velocities. With the bipolar model, we retain a warm, medium-density component in the central region, whereas with the isotropic model all the gas in the central region is dispersed and heated to high temperatures. The figure also shows the wide range of cell sizes in the simulation from $\sim50$ pc in the central region to $\sim500$ pc in the outflows entering the circumgalactic medium (CGM).
\subsubsection{Accretion model}
Our primary aim in this work is to explore how energetic the AGN feedback would need to be to significantly affect the host galaxy's evolution. To this end, we let the black hole accrete at a fixed fraction  $f_\mathrm{Edd}$ of the Eddington rate $\dot{M}_\mathrm{Edd}$ based on the initial black hole mass $M_\mathrm{BH, initial}$ motivated by the issues discussed in Section \ref{intro}. The black hole accretion rate $\dot{M}_\mathrm{BH}$ then becomes
\begin{equation} \label{eq: BHaccrate}
\dot{M}_\mathrm{BH} = f_\mathrm{Edd} \times \dot{M}_\mathrm{Edd},
\end{equation}
where
\begin{equation} \label{eq: Eddrate}
\dot{M}_\mathrm{Edd} = \frac{4\pi G M_\mathrm{BH} \mathrm{m}_\mathrm{p}}{\epsilon_\mathrm{r}\sigma_\mathrm{T} \mathrm{c}},
\end{equation}
where $G$ is the gravitational constant,  $\mathrm{m}_\mathrm{p}$ is the mass of a proton, $\epsilon_\mathrm{r}=0.1$ is the assumed radiative efficiency, $\sigma_\mathrm{T}$ is the Thomson cross-section, and $\mathrm{c}$ is the speed of light.

At each time-step $\Delta M =\dot{M}_\mathrm{BH} (1-\epsilon_\mathrm{r})\Delta t $ gets drained from the gas cells within $h_\mathrm{BH}$ using a top-hat weighting scheme so that each cell loses $\Delta m = \Delta M \times m_\mathrm{cell} / \Sigma  m_\mathrm{cell} $. We apply a limit to the mass being drained from a given cell to 90 per cent of that cell's mass. In the case of the bipolar scheme, we only drain mass from the regions outside the bipolar feedback cone.
\subsubsection{Injection model}
For isotropic feedback, the AGN luminosity is set to a fraction $\epsilon_\mathrm{r}$ of the accreted rest-mass energy. In each time-step we then inject $\Delta E_\mathrm{feed}= \epsilon_\mathrm{f} \epsilon_\mathrm{r} \Delta M \mathrm{c}^2$ where $ \epsilon_\mathrm{f}=0.05$ is the feedback efficiency. The energy is distributed over all cells within $h_\mathrm{BH}$ in an isotropic manner using a cubic spline kernel weighting. 

For non-isotropic feedback, we inject the feedback energy into a bipolar cone of opening angle $\theta_\mathrm{out} = \pi / 3$. In addition, we include the entrainment of mass into this feedback scheme by defining an efficiency parameter $\epsilon_\mathrm{out}$ which represents the fraction of gas that is swept up into the outflow. Here we set $\epsilon_\mathrm{out}=0.5$ following \citet{Curtis2015}. The luminosity then gets reduced to $\epsilon_\mathrm{r}(1-\epsilon_\mathrm{out})\dot{M}_\mathrm{BH} c^2$. Both mass and energy get injected into the cells that are within the bipolar cone and $h_\mathrm{BH}$ according to the cubic spline kernel weighting. Note that if we want to compare the two injection schemes at fixed luminosity, we need to double the accretion rate for the bipolar feedback.
\subsection{Adjustments} \label{Adjustments}
We do not want stars to form in the refinement region as having star particles of significantly lower mass colliding with more massive particles would lead to spurious N-body heating effects. \citet{Curtis2015} take care of that issue by setting the SFRs of all gas cells in the refinement region to zero. Since a central aspect of our project is to address the question of whether an AGN would be able to quench star formation, we still record the theoretical SFRs for gas cells in the refinement region but do not allow star particles to form there. 

Different from the modified equation of state models, the \citet{Smith2018} model resolves individual SN events in time and directly relates them to star particles rather than modelling SN feedback non-locally. We therefore need to include an extra mechanism to prevent SN events in the refinement region: if the mass of the gas host cell is less than 20 per cent of the target gas mass we discard the SN. This ensures that the ejecta mass will never be on the same order or larger than the host cell mass. Also since we are using the aggressive refinement, a cell's shape might not be regular for a certain fraction of time which could cause issues for the isotropic weighting scheme in the SN feedback scheme.  We keep track of the fraction of discarded SN events to ensure that we are not losing a significant amount of stellar feedback (see Appendix \ref{SuperLagRef}).

\section{Simulations} \label{section:Sims}
\subsection{Initial Conditions} \label{ICs}
The initial conditions comprise an isolated galaxy consisting of a stellar and gas disc, and a stellar bulge surrounded by hot ($10^6$ K), low-density ($10^{-6}\ \mathrm{cm^{-3}}$) gas representing the CGM. The dark matter component is modelled by a static background potential in the form of an NFW profile \citep{Navarro1997} with concentration parameter $c=10$ and spin parameter $\lambda=0.04$. The baryonic component of the initial conditions was originally generated with the \textsc{MakeDisk} code \citep{Springel2005}. Note that the star particles from the initial disc and bulge component do not contribute to the stellar feedback.

Here we focus on a system of total mass $10^{11}\ \mathrm{M_{\odot}}$, with 3.5 per cent of the mass in the disc and 0.35 per cent of the mass in bulge component. The disc is gas-rich with a gas fraction $f_\mathrm{gas}=0.5$. This allows SNe and AGN to be effective in our simulations as both require a substantial gas reservoir. Also note that our isolated galaxy is not necessarily a local dwarf, indeed with our high gas fraction, this system is more of a high-redshift analogue. The total stellar mass of our galaxy is  $2.1 \times 10^9\ \mathrm{M_{\odot}}$ which is a typical mass for observed dwarf galaxies hosting AGN \citep[see e.g.][]{Reines2015, Penny2018a}. The initial temperature of the gas disc is set to $4.6 \times 10^{4}$ K and the initial disc metallicity to $0.1\  \mathrm{Z_{\odot}}$. The virial radius of the halo is $R_\mathrm{vir}=75.5$ kpc and we choose the size of the simulated box to be $4\times R_\mathrm{vir}=302$ kpc so that the entire halo is encapsulated. The gas cells and star particles are set to have the same target mass $m_\mathrm{target}=200\ \mathrm{M_{\odot}}$. The minimum gravitational softening length for the gas cells and the fixed gravitational softening length for the star particles are set to the same value $\epsilon_\mathrm{grav}=3.8$ pc. Note that this is the same system as the `large galaxy' in \citet{Smith2018}. The only modification we make to this set-up is to add a central black hole with mass $M_\mathrm{BH}=10^5\ \mathrm{M_{\odot}}$.

We choose this black hole mass, since on the one hand it is comparable to black hole masses for systems with similar bulge masses in the literature \citep[see e.g.][]{Chilingarian2018a} and on the other hand it allows us to test AGN feedback in the IMBH range, $10^2\ \mathrm{M_{\odot}} < M_\mathrm{IMBH} < 10^5\ \mathrm{M_{\odot}}$. Note however that this black hole mass represents a conservative choice and using the $M_\mathrm{BH}-M_{\ast}$ relation from \citet{Reines2015}, we obtain a black hole mass of approximately $5\times10^5\ \mathrm{M_{\odot}}$.

\subsection{Simulation Details} \label{feed_config}
\begin{figure}
\centering
\includegraphics[width=\columnwidth]{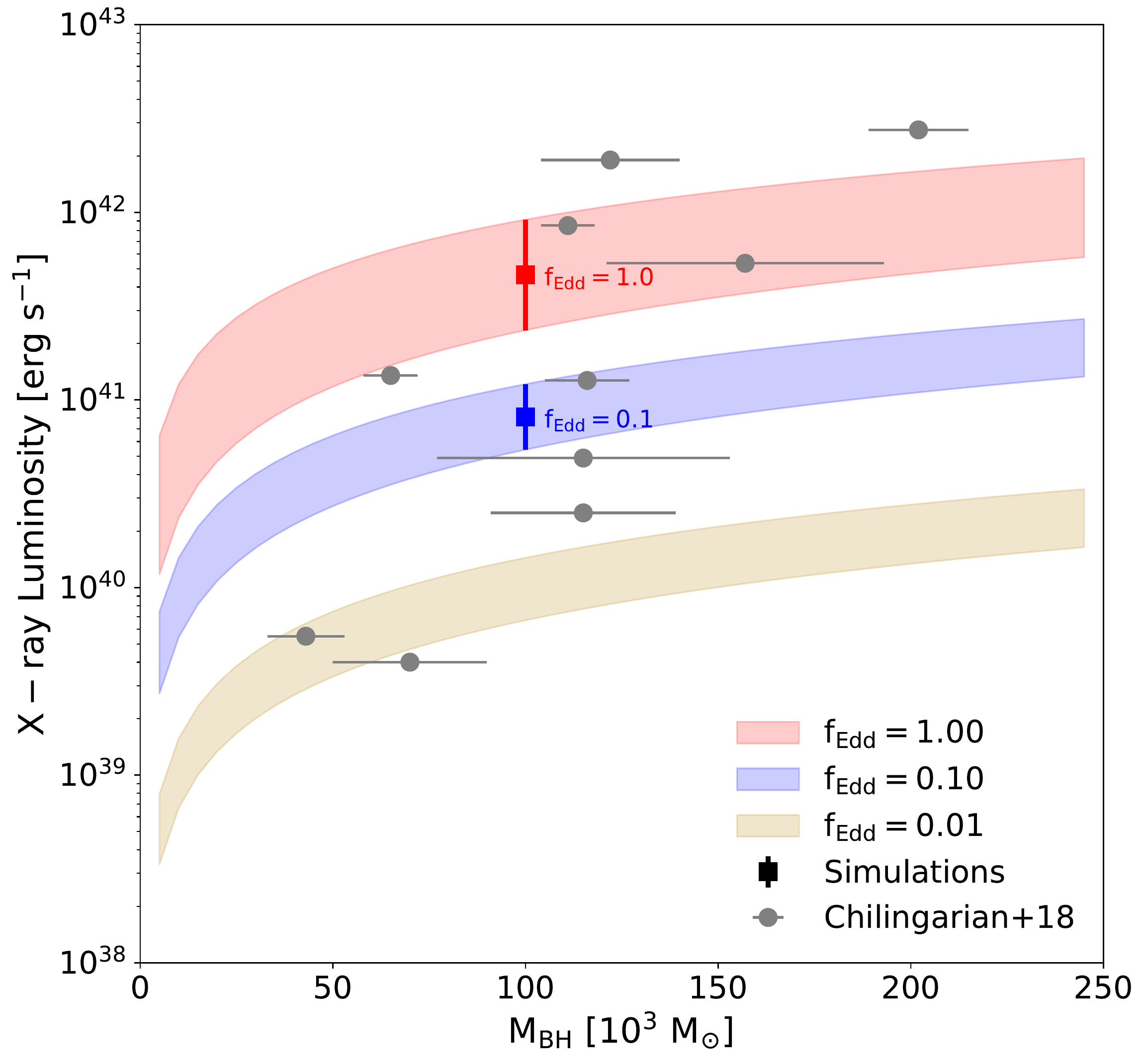}
\caption{X-ray luminosity against black hole mass. Observational data of active IMBHs from \citet{Chilingarian2018a} are shown in grey. The shaded regions show Eddington fractions $f_\mathrm{Edd}=0.01,0.1,1.0$, where the lower boundary is given by an Eddington fraction dependent \citep{Vasudevan2007} and the upper boundary by a luminosity dependent \citep{Hopkins2006} bolometric correction. Our two simulation set-ups are indicated as squares.}
\label{Xray_lum} 
\end{figure} 
We would like to identify the set-up where AGN activity has the most significant impact on dwarf galaxies. We start off with what we consider to be the most promising configuration and then investigate different variants of this fiducial run to assess which conditions are required for suppressing star formation.

Observations of intermediate mass AGN consistently find high Eddington fractions. Whilst this is obviously influenced by an observational bias, it does demonstrate that there is a non-negligible population of dwarf galaxies with AGN accreting efficiently. Figure \ref{Xray_lum} shows the observed X-ray luminosities and IMBH masses of ten confirmed intermediate mass AGN \citep{Chilingarian2018a}. For comparison, we also show the regions corresponding to different Eddington fractions on the diagram. There is a lot of uncertainty associated with converting between X-ray and bolometric luminosities. Some argue that the bolometric correction, $\kappa$, should increase with luminosity \citep[see e.g.][]{Hopkins2006}, whilst others find that Eddington fraction is a stronger discriminator between the high and low $\kappa$ populations \citep{Vasudevan2007}. In our context this difference is crucial as the AGN considered here have low luminosities, due to the low black hole mass, but high Eddington fractions. We therefore show our inferred X-ray luminosities for different Eddington fractions as shaded regions with the lower bounds calculated using the \citet{Vasudevan2007} correction and the upper bounds are obtained using the bolometric correction from \citet{Hopkins2006}.  Note that two out of ten observed AGN luminosities lie above 100 per cent of the inferred Eddington rate. However these IMBHs might have been observed at the peak of a duty cycle, and in that case using the maximum observed luminosity constantly in our simulations would likely overestimate the impact of the AGN. We therefore choose the more conservative Eddington fractions of 10 or 100 per cent for our runs (indicated as coloured squares in Figure \ref{Xray_lum}). Furthermore, we set our fiducial instantaneous star formation efficiency to $\epsilon_\mathrm{SF}=0.15$ to obtain high SN efficiency.

We hypothesise that the isotropic thermal feedback would be most effective in influencing star formation, as it drives energy directly into the disc. We therefore choose the run with an AGN emitting thermal energy isotropically at 100 per cent of the Eddington rate as our fiducial set-up (SN+AGNTh100). We also repeat the same run with the AGN shining at 10 per cent of the Eddington luminosity (SN+AGNTh10), and with the AGN shining at 100 per cent of the Eddington luminosity, but with the energy input being restricted to a bipolar cone and additional mass loading (SN+AGNBi100). Note that we expect the latter to be less efficient at regulating star formation as most of the thermal energy will simply escape from the galaxy. Finally, we also test a set-up with a lower star formation efficiency of  $ \epsilon = 0.015 $ (SN+AGNTh100Low). 

For comparison we also carried out SN only runs as well as runs without any feedback at both star formation efficiencies (SN, SNLow, NoFeedback, and NoFeedbackLow). All simulation runs and their properties are listed in Table \ref{runs}. 

\begin{table}
\caption{Overview of the simulation runs, where we list star formation efficiencies ($\epsilon_\mathrm{SF}$), AGN luminosities, and ``switched-on" feedback mechanisms.}
\centering
\begin{tabular*}{\columnwidth}{@{}lllll@{}}
\toprule
Name & $\epsilon_\mathrm{SF}$  & $\frac{L_\mathrm{AGN}}{L_\mathrm{Edd}}$ & Feedback Physics \\ \midrule
NoFeedback & 0.15                 & 0.0      & no energetic feedback                            \\
$\mathrm{NoFeedbackLow}$ & 0.015                 & 0.0      & no energetic feedback                            \\
SN & 0.15  			               & 0.0     & SNe                                  \\
$\mathrm{SNLow}$ & 0.015  			               & 0.0     & SNe                                   \\
SN+AGNTh100 & 0.15   			                 & 1.0 & SNe,  AGN (thermal)                                 \\
$\mathrm{SN+AGNTh100Low}$ & 0.015   			                 & 1.0 & SNe,  AGN (thermal)                                   \\
SN+AGNTh10 & 0.15   			                 & 0.1 & SNe,  AGN (thermal)                                   \\
SN+AGNBi100 & 0.15   			                 & 1.0 & SNe,  AGN (bipolar)                                   \\ \bottomrule
\end{tabular*}
\label{runs}
\end{table}

\section{Results}\label{section:Results}

\subsection{Gas properties}

\begin{figure*}
\includegraphics[width=\textwidth]{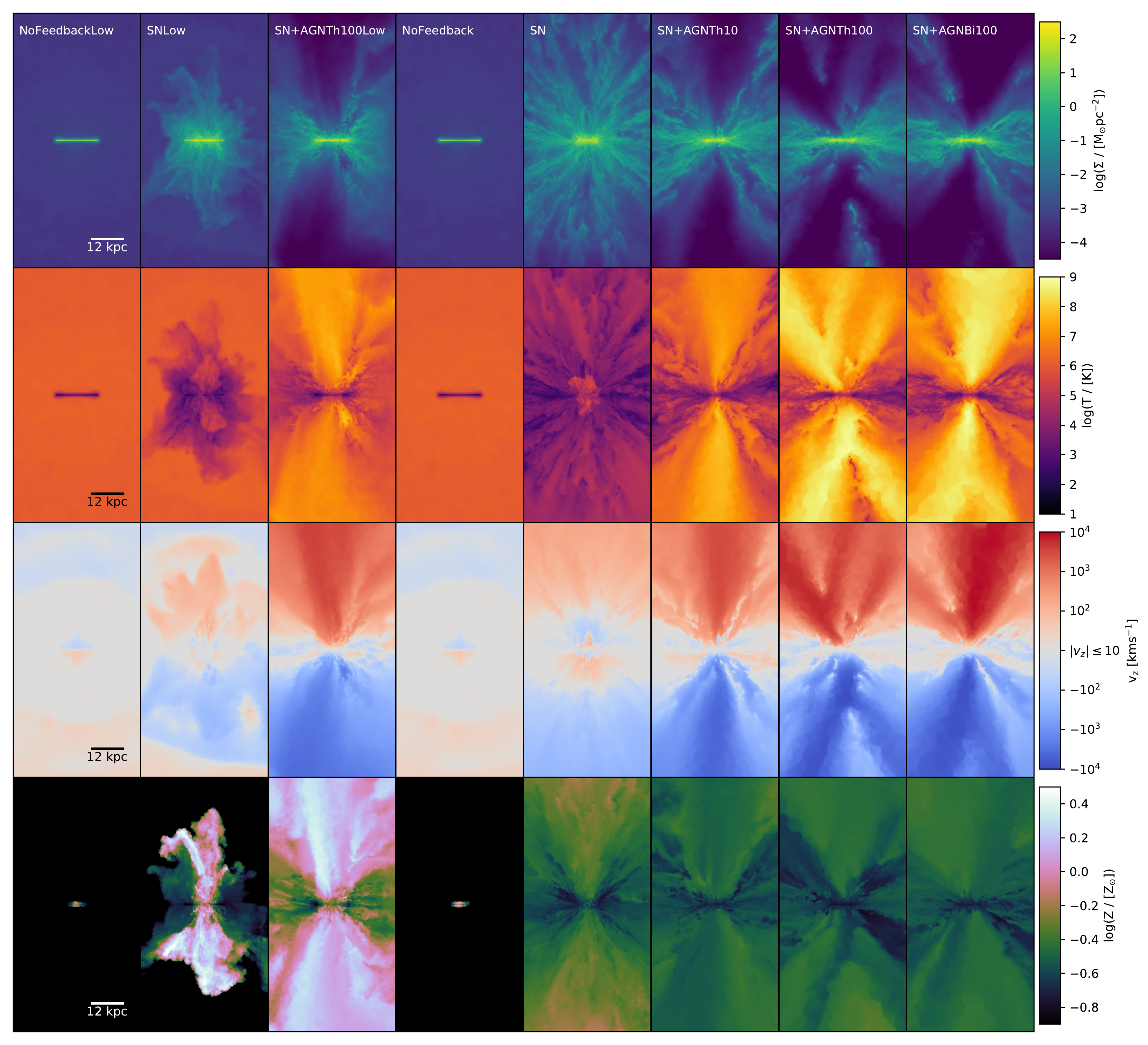}
\caption{$50 \times 100 \times 10$ kpc edge-on projections of all the simulation runs at $t=300$ Myr. The first row shows surface densities, the second row shows temperatures, the third row shows vertical velocities, and the fourth row shows metallicities. The low $\epsilon_\mathrm{SF}$ runs are depicted on the left hand side, and the high $\epsilon_\mathrm{SF}$ runs on the right hand side. With only SN feedback, the outflows are slow, warm, and high-density. With additional AGN feedback, the outflows have an additional hot, low-density component moving at high velocities. The differences in outflow metallicities at $t=300$ Myr are largely driven by differences in outflow histories rather than star formation. }
\label{fig:all_proj} 
\end{figure*} 

We start our analysis by inspecting the visuals of the simulation suite which are shown in Figure \ref{fig:all_proj}. Here, $50 \times 100 \times 10$ kpc edge-on projections of surface density, temperature, vertical velocity and metallicity at $t=300$ Myr are plotted. As expected, both of the runs without energetic feedback (`NoFeedbackLow' and `NoFeedback') overcool resulting in a massive overproduction of stars. The high $\epsilon_\mathrm{SF}$ run has slightly more metals as even more stars are formed with this set-up. None of the two no feedback runs have outflows.

Adding SN feedback to this set-up triggers large-scale outflows. However at low $\epsilon_\mathrm{SF}$, the outflows are much weaker consisting of mainly slow-moving, relatively dense gas and only reaching heights of  $z\sim 35$ kpc. The high $\epsilon_\mathrm{SF}$ run drives much stronger outflows which reach heights $z>50$ kpc. The outflows have a significant warm component ($T\sim10^{5}$ K), but are still relatively slow ($v_\mathrm{z}\sim10^2\ \mathrm{kms^{-1}}$). Close to the disc, both SN runs have small-scale hot outflows, reaching temperatures $T\sim10^{6}$ K. We also observe a significant galactic fountain effect close to the disc which is even stronger for the high $\epsilon_\mathrm{SF}$ run due to a recent quenching episode which weakened the SN feedback. 

With additional AGN activity outflows are significantly enhanced. At low $\epsilon_\mathrm{SF}$, the outflows reach extremely high velocities of several thousands $\mathrm{kms^{-1}}$. The bulk component of the outflowing material is hot ($T\sim 10^{7}\ \mathrm{K}$), low-density, and high-metallicity ($Z \gtrsim \mathrm{Z_{\odot}}$). While in the `SNLow' run, the propagation of metals into the CGM is limited, the outflows in the `SN+AGNTh100Low' run are powerful enough to increase the metallicity in the whole surrounding region. 

At high $\epsilon_\mathrm{SF}$, the AGN runs have lower metallicity than the `SN' run within the extent of the projections shown. Though this difference is not driven by SN activity, but by different outflow properties\footnote{In the AGN runs, the metals from the initial starburst get ejected earlier as the AGN triggers outflows from the very beginning, different from the SN feedback which takes $\sim 40$ Myr to develop its full effect. Furthermore, the AGN runs eject material at a higher rate, so at $t=300$ Myr, the high metallicity material has reached heights $z>50$ kpc.}. Suppression of central SFRs by the AGN is a secondary effect, the overall metal mass between `SN' and `SN+AGNTh100' differs by only three per cent.

The AGN runs at high  $\epsilon_\mathrm{SF}$ are significantly more multi-phase than the other set-ups. Outflow velocities range from tens to thousands $\mathrm{kms^{-1}}$ and outflow temperatures from $\sim10^{4}$ K to $\gtrsim10^{9}$ K. As the simulation evolves the hot gas fraction increases due to the constant input of thermal energy.

\begin{figure*}
\centering
\includegraphics[width=\textwidth]{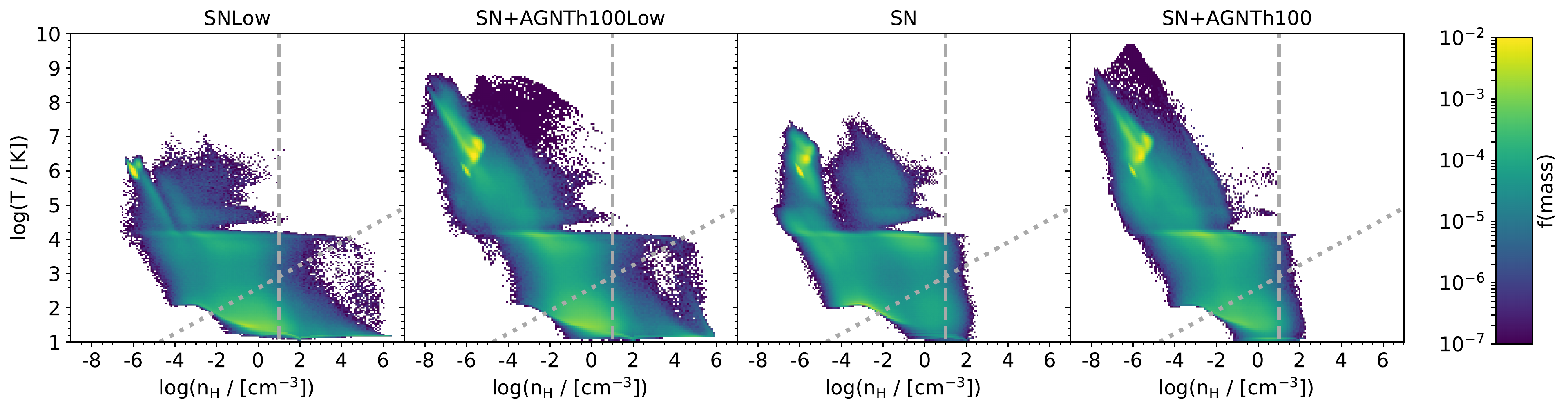}
\caption{Comparison of the phase diagrams for the two star formation efficiencies at $t=300$ Myr.  The models used are labelled on the panels. Colour coding corresponds to the fraction of gas mass within a given pixel. The dashed line indicates the star formation threshold and the dotted line indicates the non-thermal pressure floor. For the runs where feedback is efficient (`SN+AGNTh100Low', `SN' and `SN+AGNTh100'), a significant amount of mass is located in the low-density, high-temperature region of the phase diagram, representing hot outflows. For the AGN runs, these outflows reach extremely high temperatures up to $T\sim10^9$ K. The gas above the star formation density threshold is barely affected by the addition of AGN feedback.}
\label{fig:phase_diagram}
\end{figure*}

We can compare the effect of different energy injection schemes and different $\epsilon_\mathrm{SF}$ in more detail by looking at the phase diagrams. Figure \ref{fig:phase_diagram} shows the phase diagrams of the gas for the SN and high luminosity thermal wind runs at $t=300$ Myr. Colour coding represents the fraction of mass in a given pixel. For all runs, the CGM gas is visible at $T\sim10^6$ K and $n_\mathrm{H}\sim 10^{-6}\ \mathrm{cm^{-3}}$ and the ISM gas at $T \lesssim  4.5\times10^{4}$ K. The connection between those two regions represents gas cooling from the CGM onto the disc.  The dotted line indicates the non-thermal pressure floor and the star formation density threshold is shown as a vertical dashed grey line.

The low $\epsilon_\mathrm{SF}$ runs all have significant amounts of dense, cool gas above this threshold  ($\sim 5$ per cent of gas is star-forming in both cases). The gas has short cooling times and mostly occupies the high-density, low-temperature region of the phase diagram. Nevertheless, some of the gas manages to get to higher temperatures, where cooling times are longer. We obtain small-scale outflows for the `SNLow' run, with some gas ($M_\mathrm{hot}\sim9\times10^{2}\ \mathrm{M_{\odot}}$) reaching temperatures above $\mathrm{10^{7}}$ K. The `SN+AGNTh100Low' run produces large outflows with significant amount of gas in the hot phase ($M_\mathrm{hot}\sim7\times10^{7}\ \mathrm{M_{\odot}}$).  For the $\epsilon_\mathrm{SF} = 15$ per cent runs, we have strong outflows both with and without an AGN. The SN feedback drives powerful outflows ($M_\mathrm{hot}\sim3\times10^{6}\ \mathrm{M_{\odot}}$) and efficiently regulates star formation. The AGN activity increases the hot mass by approximately two orders of magnitude to $M_\mathrm{hot}\sim1\times10^{8}\ \mathrm{M_{\odot}}$, and the outflows are more multiphase also containing warm components (cf. Figure \ref{fig:all_proj}). As with the low $\epsilon_\mathrm{SF}$ case, the gas above the star formation density threshold is barely affected by the AGN and for both high $\epsilon_\mathrm{SF}$ runs, we have a star-forming gas mass fraction of less than one percent.

For the runs where feedback is efficient (`SN+AGNTh100Low', `SN' and `SN+AGNTh100'),  a significant amount of mass is located in the low-density, high-temperature region of the phase diagram at $\mathrm{10^{-6}\ cm^{-3}\lesssim n_{H} \lesssim 10^{-5}\ cm^{-3}}$ and $ \mathrm{10^{6}\ K \lesssim T \lesssim 10^{7}}$ K. This feature represents the hot outflows, which carry a significant amount of gas mass away from the galaxy (see Section \ref{section:outflows}). For the two AGN runs, we also have a second high-temperature feature. As can be seen from the colour coding, these cells are super-refined gas cells within the black hole smoothing length. This feature is then explained as a by-product of the super-Lagrangian refinement scheme together with a minimum mass constraint\footnote{As discussed in Appendix \ref{SuperLagRef}, a significant number of cells will reach the minimum mass which is set to $2\times10^{-1}\ \mathrm{M}_\mathrm{\odot}$.  This leads to the strip in the phase diagram of the `SN+AGNTh100' run. In the low  $\epsilon_\mathrm{SF}$ case, higher central surface densities mean that fewer cells will reach the minimum mass and therefore the distribution of the gas phases in the refinement region is broadened.}.

\subsection{Star formation} \label{section:sf}
\begin{figure*}
\centering
\includegraphics[width=0.49\textwidth]{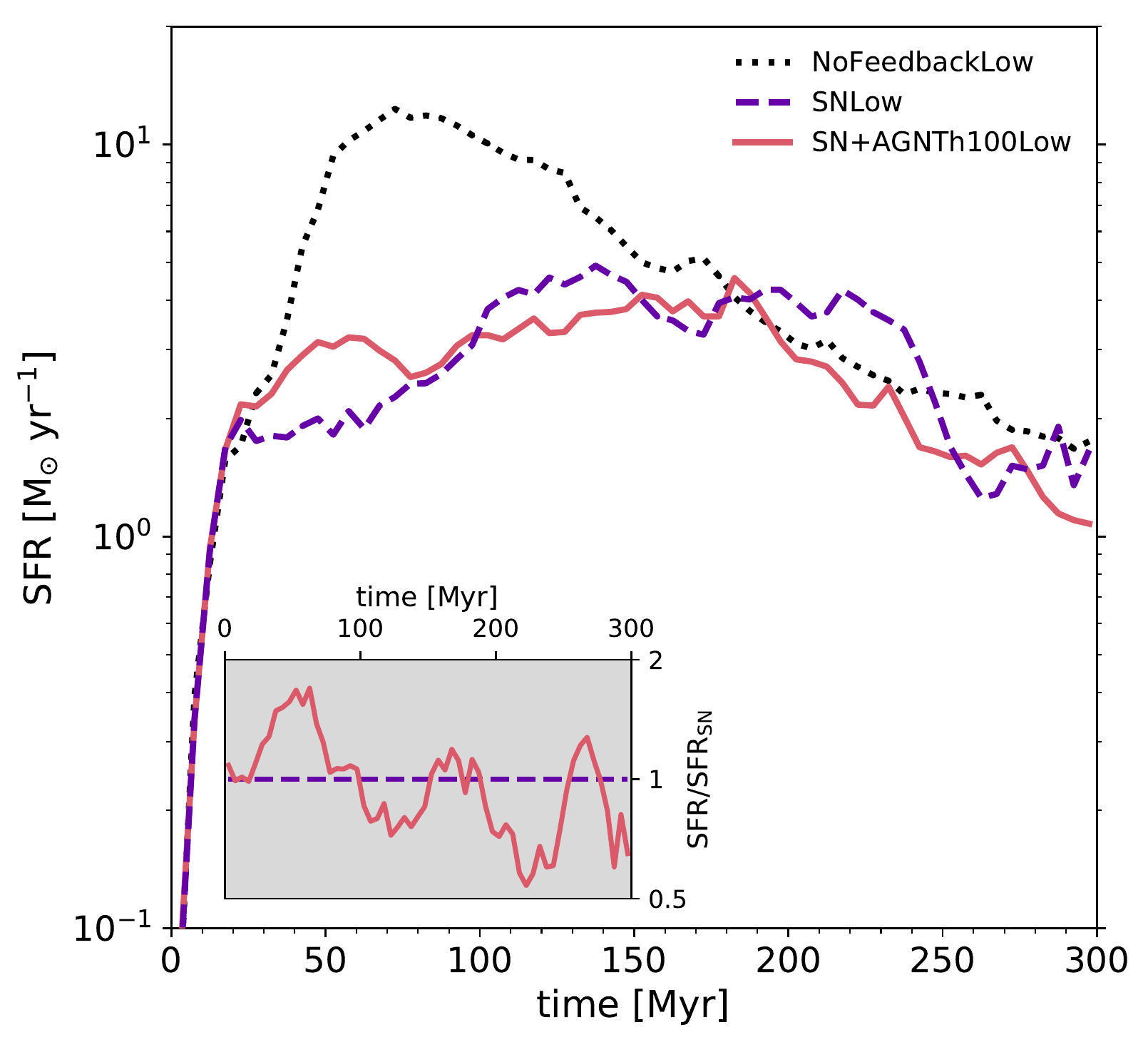}
\includegraphics[width=0.49\textwidth]{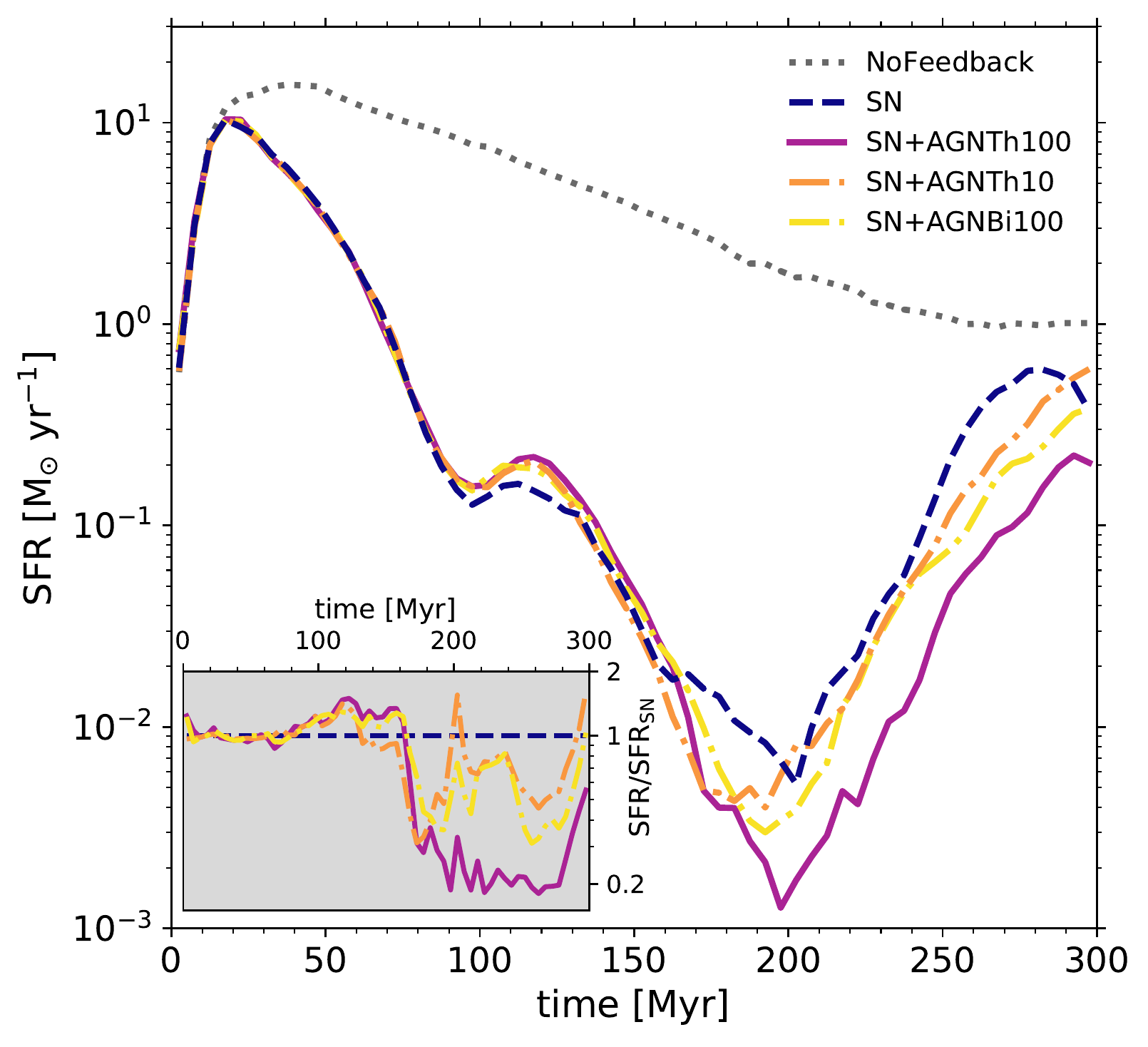}
\caption{\textit{Left panel:} SFRs for the  $ \epsilon_\mathrm{SF} = 1.5$ per cent runs. The inset shows the ratio between the AGN and SN only runs. Adding AGN outflows does not have a significant effect on SFRs in this set-up. \textit{Right panel:} SFRs for the  $ \epsilon_\mathrm{SF} = 15$ per cent runs. The inset shows the ratio between the AGN and SN only runs. All AGN outflow models affect SFRs at late times. The high-luminosity, thermal run (`SN+AGNTh100') is the most effective, suppressing the SFR by approximately a factor of five for $t\gtrsim 200$ Myr.}
\label{fig:sfr}
\end{figure*}
\begin{figure*}
\centering
\includegraphics[width=\textwidth]{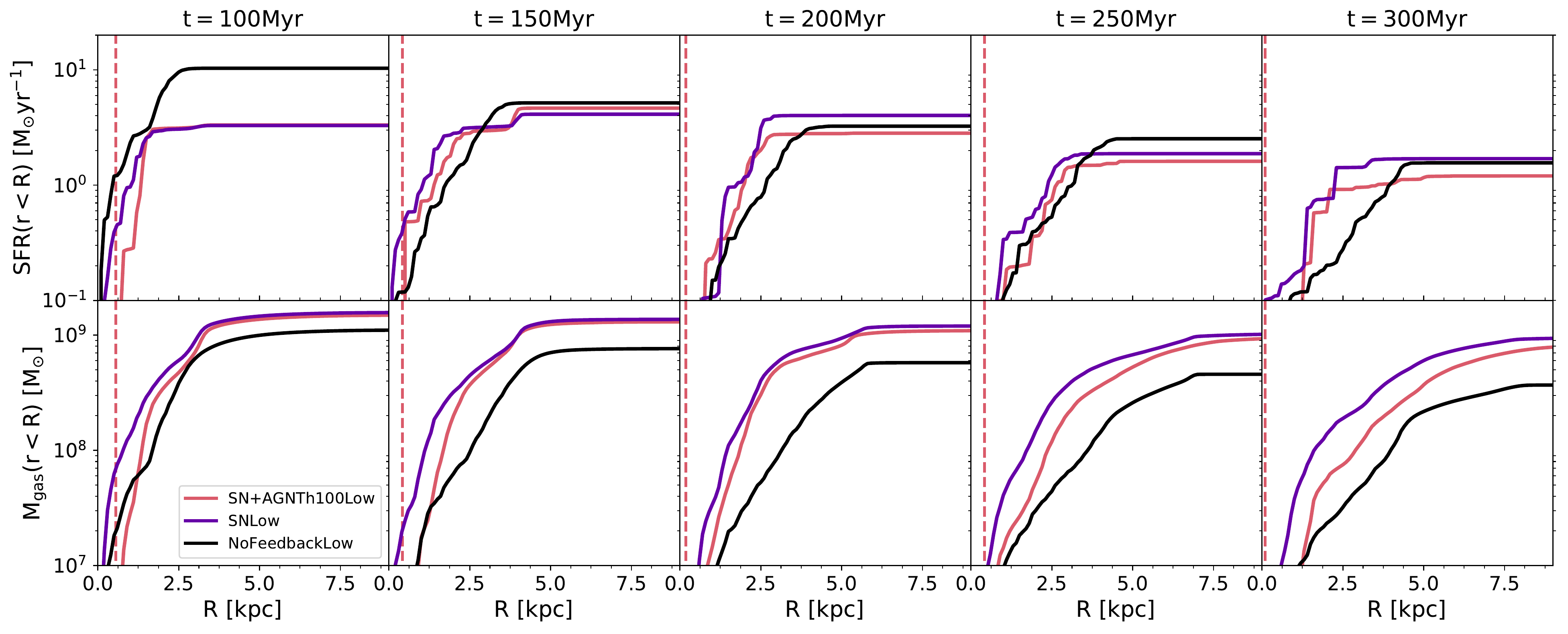}\\
\vspace{2ex}
\includegraphics[width=\textwidth]{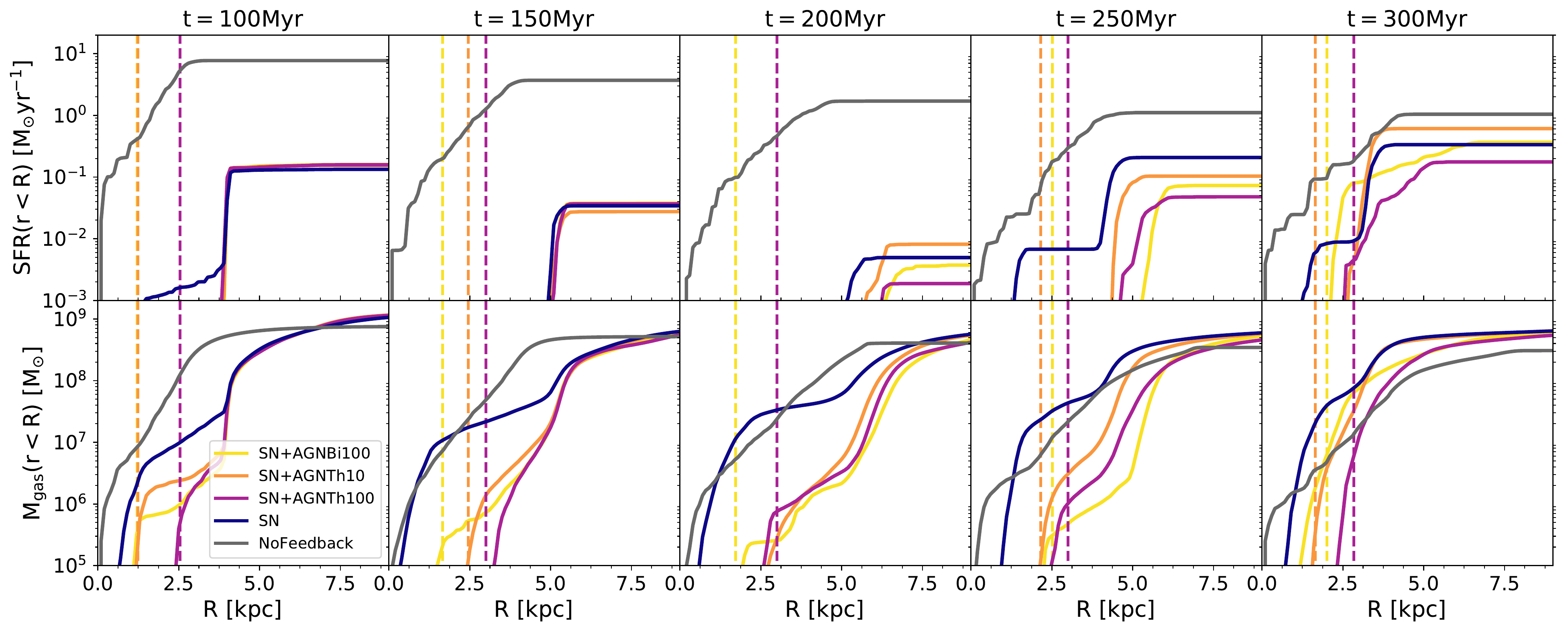}
\caption{Cumulative SFRs as a function of distance from the galaxy centre for $\epsilon_\mathrm{SF} = 1.5$ per cent (top panel) and $\epsilon_\mathrm{SF} = 15$ per cent (bottom panel) runs. For comparison cumulative gas masses are also shown in each case. The black hole smoothing lengths (equal to the size of the refinement region) are indicated as vertical dashed lines. For the runs with AGN feedback, central SFRs are systematically suppressed, in particular at high $\epsilon_\mathrm{SF}$.}
\label{SFR_vs_r}
\end{figure*}
Before we delve into the details of outflows in section \ref{section:outflows}, we want to assess the impact of AGN activity on star formation which is one of the most fundamental quantities of galaxy formation. Figure \ref{fig:sfr} shows the SFRs for  $ \epsilon_\mathrm{SF} = 1.5$ per cent and $ \epsilon_\mathrm{SF} = 15$ per cent. For the $ \epsilon_\mathrm{SF} = 1.5$ per cent case, the no feedback run has a starburst between $\sim30$ and 150 Myr using up most of the gas and then follows the SFRs of the two feedback runs. The `SN+AGNTh100Low' run forms slightly more stars than the `SNLow' run early on\footnote{This a positive feedback effect from gas being compressed on the expanding AGN bubble.} and then has roughly similar, albeit systematically lower, SFRs. This can be seen more clearly in the inset which shows the ratio between the AGN and the SN only run.

For the  $\epsilon_\mathrm{SF} = 15$ per cent runs, there is an early peak in the SFR at $t \sim 25$ Myr. Afterwards, the SFR of the `NoFeedback' run slightly decreases, with the majority of the high-density gas having been consumed, while the SFRs of the runs including SN feedback decrease rapidly. This makes sense as a burst of star formation will be followed by a burst of SN activity. At around $t\sim90$ Myr, the SFR of the `SN' run' drops slightly below the SFR of the AGN runs. Finally, at around $t \sim 160$ Myr, the AGN outflows start taking effect. The SFRs of all three AGN runs start falling below the `SN' run. The `SN+AGNTh100' run has the most significant effect, decreasing SFRs by by approximately a factor of five for $t\gtrsim 200$ Myr. The other two AGN runs have qualitatively similar effects, though the offset is less significant here. Since the bipolar mode only injects energy into a double cone perpendicular to the disc, we would not expect it to have a large effect on the cold disc gas - the isotropic thermal winds can have a much more significant impact here. Also note that towards the end of the simulation at $t\sim300$ Myr, the SFRs start decreasing again. We followed the `SN' and `SN+AGNTh100' slightly longer and found that the pattern of SFR peak followed by SFR trough (due to high SN activity) keeps repeating, as one would expect with a strong stellar feedback loop. 

From the temporal evolution of star formation properties, one could conclude that AGN outflows do not significantly affect star formation in dwarf galaxies. However, looking at the spatial distribution instead, we see a small but systematic offset with star formation being suppressed in the central region for different AGN models, luminosities and star formation efficiencies. Figure \ref{SFR_vs_r} shows the SFR within a given radius $R$ for $\epsilon_\mathrm{SF} = 1.5$ per cent and $ \epsilon_\mathrm{SF} = 15$ per cent. For reference, the enclosed gas mass is also shown in the lower panels. The black hole smoothing lengths are indicated as vertical dashed lines. 

For the low $\epsilon_\mathrm{SF}$ runs,  central SFRs are slightly suppressed when including AGN outflows on top of SNe at $t=100$, 250, 300 Myr. For the high $\epsilon_\mathrm{SF}$ runs, this effect is more pronounced. Here the AGN runs have lower central SFRs for the majority of the simulation times. In all cases, the AGN activity has significantly depleted the central region, though note that at fixed luminosity the bipolar run has a much smaller effect on the gas distribution than the isotropic thermal run. This is largely due to energy injection rather than mass accretion effects, as the accreted mass over $t=300$ Myr amounts to merely $M_\mathrm{acc,tot}=f_\mathrm{Edd} \times 6.7 \times  10^{5}\ \mathrm{M_{\odot}}$, whilst the offsets observed are at least on the order of $10^{6}-10^{7}\ \mathrm{M_{\odot}}$.

Overall, we find that, even with an AGN shining at 100 per cent of the Eddington luminosity, we cannot suppress star formation globally, though we do observe significant effects locally. Partly this is due to the SN feedback being very efficient so that majority of the quenching work has already been done. Another reason that AGN activity does not affect star formation is that most of the AGN feedback energy escapes along the path of least resistance, i.e. in this case perpendicular to the disc \citep[see e.g.][]{Wagner2013,Bourne2014,Costa2014,Gabor2014,Bieri2017}.

In addition there are some caveats with our isolated model, the most important one being that we do not have gas inflows. If we had gas inflows these would likely be hindered by an AGN, as we see significant global effects of AGN activity on galactic outflows (cf. Figure \ref{fig:all_proj}). These effects will be analysed in more detail in the next section.
\subsection{Outflow properties} \label{section:outflows}

\begin{figure*}
\centering
\includegraphics[width=\textwidth]{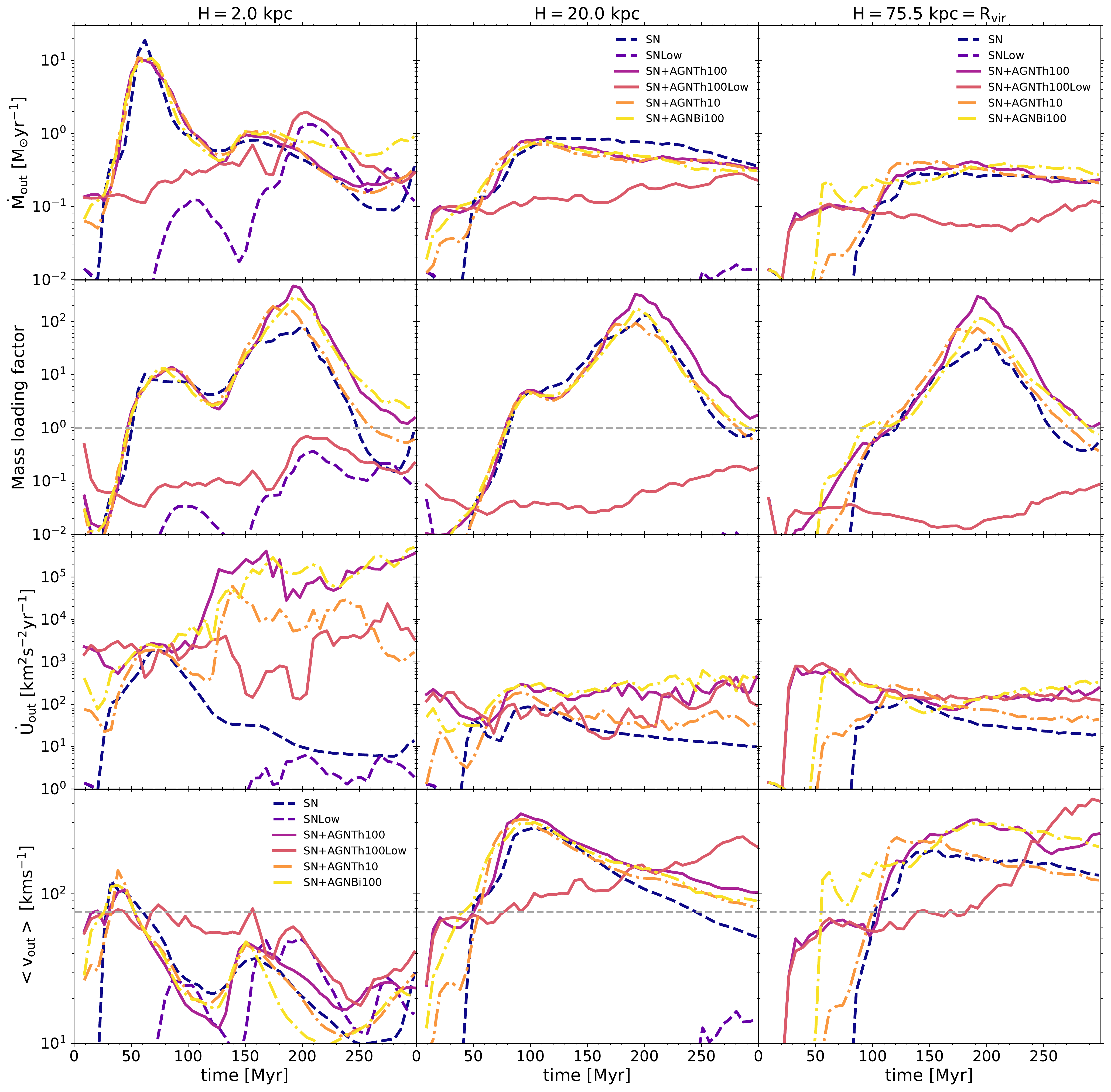}
\caption{Outflow properties against time for the  $\epsilon_\mathrm{SF} = 1.5$ per cent and the  $\epsilon_\mathrm{SF} = 15$ per cent runs calculated at target heights $\mathrm{H=2\ kpc}$  (left column),  $\mathrm{H=20\ kpc}$ (middle column), and $\mathrm{H=75.5\ kpc=R_\mathrm{vir}}$ (right column). The first row shows the mass outflow rates $\dot{M}_\mathrm{out}$, the second row shows the mass loading factor, the third row shows the thermal energy outflow rates $\dot{U}_\mathrm{out}$, and the fourth row shows the average outflow velocities $<v_\mathrm{out}>$ (the virial velocity is indicated as a dashed grey line). The different models used are given by the figure legend. The `SNLow' run, in general, does not have powerful enough feedback to drive significant outflows, and even by adding an AGN shining at 100 per cent of the Eddington luminosity, we do not obtain significant mass loading factors. When SN feedback is efficient (high $\epsilon_\mathrm{SF}$ runs), the most striking difference are the thermal outflow rates, with the AGN expelling up to five orders of magnitude more thermal energy from the galaxy. }
\label{fig:outflowrates}
\end{figure*}

\begin{figure*}
\includegraphics[width=\textwidth]{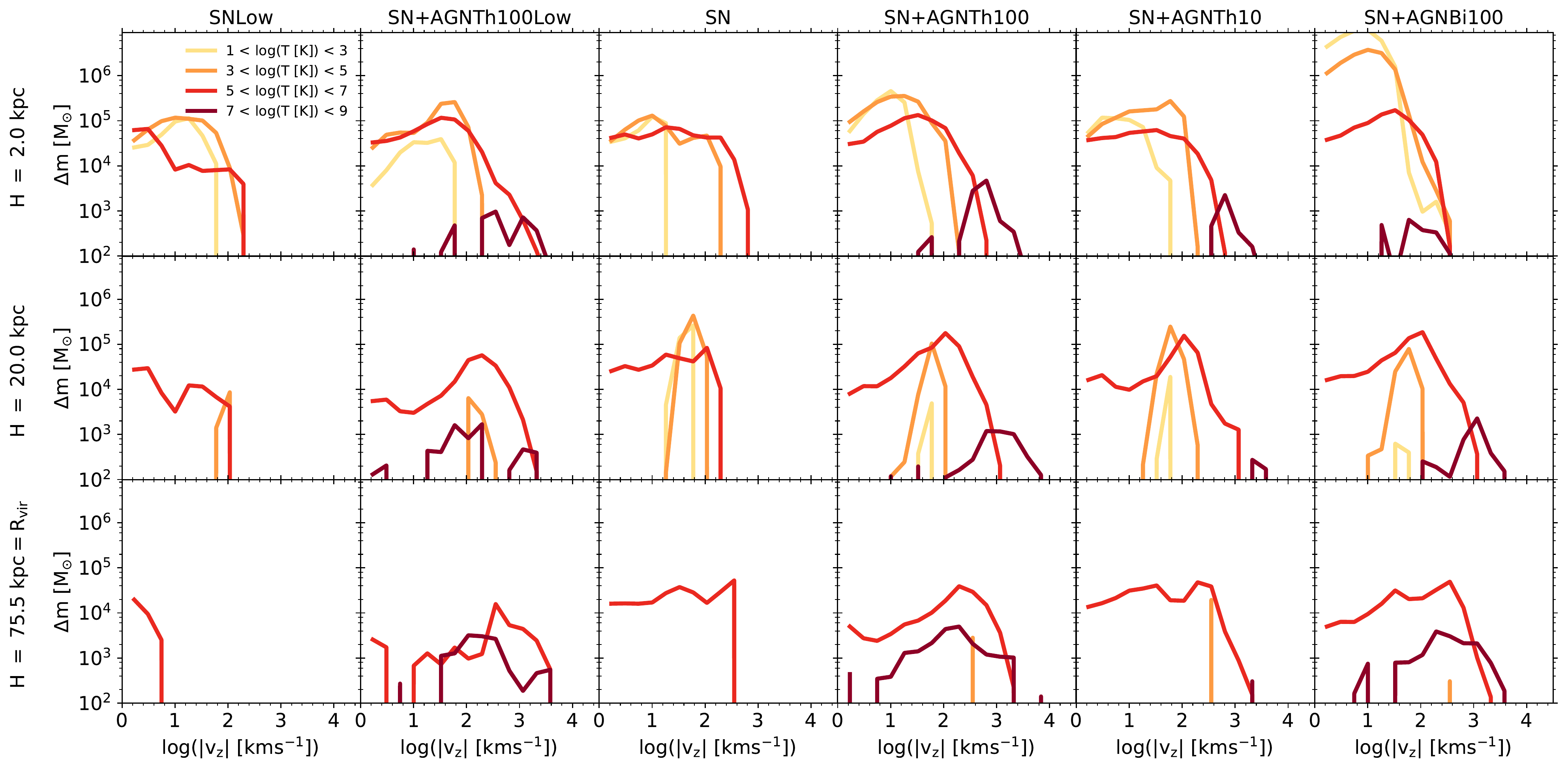}
\caption{Mass histogram of outflow velocities split by temperatures at three target heights $\mathrm{H=2\ kpc}$  (top row), $\mathrm{H=20\ kpc}$ (middle row), and $\mathrm{H=75.5\ kpc=R_\mathrm{vir}}$ (bottom row) at $t=300$ Myr. The thermal and kinematic properties of the AGN feedback runs are significantly different from their SN only equivalents even with a more conservative luminosity of ten per cent $L_\mathrm{Edd}$. Only the AGN-driven outflows carry a `very hot' outflow component with temperatures above $10^{7}$ K, and these outflows are much faster reaching several thousands $\mathrm{kms^{-1}}$.}
\label{fig:vel_temp_hist}
\end{figure*}

We further investigate the outflows visible in Figure \ref{fig:all_proj} by plotting the mass outflow rate $\dot{M}_\mathrm{out}$, the mass loading factor (ratio of mass outflow rate to SFR), the thermal energy outflow rate $\dot{U}_\mathrm{out}$ and the mass-weighted average outflow velocity $<v_\mathrm{out}>$ at three target heights ($H=2$ kpc, $H=20$ kpc, and $H=75.5\ \mathrm{kpc}=R_\mathrm{vir}$) for a slice of thickness $\mathrm{d}z=200$ pc in Figure \ref{fig:outflowrates}. The virial velocity is indicated as a dashed grey line in the panels showing the average outflow velocities. Outflow velocities are taken vertically away from the disc plane, and we only include gas flowing away from the disc by requiring $\text{sgn(}v_{z}\text{)}=\text{sgn(}z\text{)}$. 

At low $\epsilon_\mathrm{SF}$, we can see clear differences between the `SNLow' and `SN+AGNTh100Low' runs for all quantities considered. The AGN runs have large-scale outflows from the beginning of the simulation run. The SN only runs have small-scale outflows from $t\sim50$ Myr. This is because it takes some time for stars to form and explode as SNe, whilst the AGN is active from the beginning of the simulation. The SN-driven outflows at $H=2$ kpc reach mass outflow rates similar to the `SN+AGNTh100Low' run. The outflows at $H=20$ kpc, however, are about one order of magnitude weaker and only reach significant outflow rates from $t=250$ Myr onwards. For `SNLow', no outflows reach the virial radius, within the simulation time. This shows that the outflows at $H=2$ kpc mainly originate from a warm fountain, which does not propagate into the CGM.

Due to the similar SFRs, the mass loading factors follow the same pattern. Note that from observations and theory, we would expect mass loading factors between one and ten, whilst we get a mass loading factor of order 0.1 for both runs at $H=2$ kpc. Remarkably, even with a sustained high-luminosity AGN, we do not obtain high mass loading. 

We also consider the outflow rate of thermal energy per unit mass. At $H=2$ kpc, there is some thermal energy outflow for the `SNLow' run. This is mainly associated with the warm fountain, as this component is not present at larger heights. The `SN+AGNTh100Low' runs has significant thermal energy outflow rates, comparable with the high $\epsilon_\mathrm{SF}$ AGN runs. 

Finally, we compare the mass-weighted average outflow velocities. Note that these will be biased towards lower velocities, as we also capture the slow-moving CGM gas which happens to have its (extremely small) velocity orientated away from the disc. Close to the disc, at $H=2$ kpc, the `SN+AGNTh100Low' outflow velocities are initially quite high at $60-80\ \mathrm{kms^{-1}}$, but then decrease at later times as the additional slow-moving SN-driven outflow lowers the average velocity. At the virial radius there is a sharp increase in outflow velocities from $t\sim200$ Myr onwards as there is no more SN-driven component, and all of the material is hot. Also note that further away from the disc, at $H=20, 75.5$ kpc, the average outflow velocity is consistently above the virial velocity at late times, suggesting that outflows might become unbound. The `SNLow' outflow velocities fluctuate between 20 and 50 $\mathrm{kms^{-1}}$ at $H=2$ kpc (in accordance with the timings of SN bursts). There are no outflows with significant velocities further away from the disc. 

For the high $\epsilon_\mathrm{SF}$ runs, we also have AGN-driven outflows from the very beginning of the simulations, whilst there is a time delay of $t\sim25$ Myr for the `SN' run, again set by the time it takes for stars to form and explode as SNe. The time delay is shorter here as the star formation is more efficient. For all runs, there is a high amount of mass outflow at $H=2$ kpc at $t\sim60$ Myr. This is caused by the extreme SN activity following the starburst at $t\sim25$ Myr (cf. Figure \ref{fig:sfr}).

Overall, the outflow rates are roughly similar for the different feedback models with slightly higher mass outflow rates for the `SN+AGNBi100' run, which makes sense as these outflows have additional mass-loading from the AGN. At large scales the outflow rates are much steadier and there is not a significant difference between the models, once SN feedback starts driving outflows. As SFRs do not differ too much between the different set-ups, the time evolution of the mass loading factors follows a similar pattern, spanning values from $10^{-2}$ to several hundreds. The first maximum at $H=2$ kpc is largely due to the initial starburst, and is absent at the other two target heights, again suggesting that the initial burst ends up as a warm fountain and the outflow does not manage to fully escape. 

Note that close to the disc, the average outflow velocity stays below the virial velocity for all times except the initial starburst. Similarly, the peak at $H=20$ kpc is also associated with the peak in SN activity. However, at the virial radius this peak vanishes, and from $t=100$ Myr onwards, all high $\epsilon_\mathrm{SF}$ runs have outflows steadily above the virial velocity, as only the most energetic gas manages to escape to this height. Note that with AGN activity the outflows at the virial radius are much faster at late times. This can be explained by noting that AGN-driven outflows are mainly low-density and hot, while the SN-driven outflows carry significant amounts of warm and relatively dense material, moving at significantly lower velocities (cf. vertical velocity projections in Figure \ref{fig:all_proj}).

This point is strengthened by the significant differences in the thermal energy outflow rates. At small scales the difference reaches more than five orders of magnitudes at $t\sim300$ Myr when SN activity is significantly suppressed due to the minimum in star formation at $t\sim200$ Myr. At large scales, we also observe significant offsets up to almost two orders of magnitude. The difference between the thermal properties decrease as the cold and warm component of the SN outflows do not make it to large disc heights so that the thermal properties far away from the disc are relatively similar.

We further investigate the kinematic and thermal properties of the different runs by inspecting the velocity distribution of the different gas phases. Figure \ref{fig:vel_temp_hist} shows the mass in different velocity bins split by temperatures (cold: $10-10^{3}$ K, warm: $10^{3}-10^{5}$ K, hot: $10^{5}-10^{7}$ K, very hot: $10^{7}-10^{9}$ K) at the three target heights $H=2$ kpc, $H=20$ kpc, and $H=75.5$ kpc at $t=300$ Myr. We also include inflows here as we are mainly interested in whether the gas is slow or fast moving rather than the direction. The CGM is clearly identifiable as the hot, slow-moving component.

The first interesting point to note is that neither the `SNLow' nor the `SN' run has a very hot component. Close to the disc, both runs have a warm and a hot component with velocities up to several hundreds $\mathrm{kms^{-1}}$, and a slow-moving cold outflow component. The `SN' run also has a significant cold component at intermediate heights ($H=20$ kpc), whilst the outflow `SNLow' run is mainly just in the hot phase. At the virial radius, the `SN' outflow has all its gas mass in the hot phase, whilst for the `SNLow' run, we are left with only the CGM component, since the outflows do not propagate far enough.

All of the AGN models result in outflows with a very hot component, and also have broader velocity distributions with outflow speeds reaching ten thousands of $\mathrm{kms^{-1}}$. Another interesting thing to note is that while all AGN outflows have a significant cold component at small scales, this is mostly lost at large scales due to lower outflow velocities in the cold phase and mixing with the very hot component. At the virial radius, all outflows are almost exclusively hot or very hot.

In the whole simulation volume, the `SN' run and the runs with AGN outflows have similar amounts of fast-moving gas ($|v_\mathrm{z}|>10^{2}\ \mathrm{kms^{-1}}$)  with $M_\mathrm{fast}\sim 10^{8}\ \mathrm{M_{\odot}}$, but while for the AGN runs 10 per cent of that fast-moving gas is very hot, for the `SN' run it is only 1 per cent. For `SNLow', we only  have $M_\mathrm{fast}\sim 10^{6}\ \mathrm{M_{\odot}}$, of which just 0.01 per cent is hot. Note that the thermal outflow properties may be resolution dependent \citep[see e.g.][]{Smith2018b,Hu2018}, though this would be a relative effect, with both SNe and AGN becoming more effective in heating the gas at higher resolution.

Overall, we conclude that the thermal and kinematic properties of the AGN runs are significantly different from their SN only equivalents even with a more conservative luminosity of ten per cent $L_\mathrm{Edd}$. These distinct differences between the kinematic properties of the simulations are promising as they carry specific observational signatures. In the next section, we discuss these possible signatures in more detail and how the results from our simulations compare to observations from MaNGA.

\subsection{Comparison to MaNGA}
\begin{figure*}
\centering
\includegraphics[width=\textwidth]{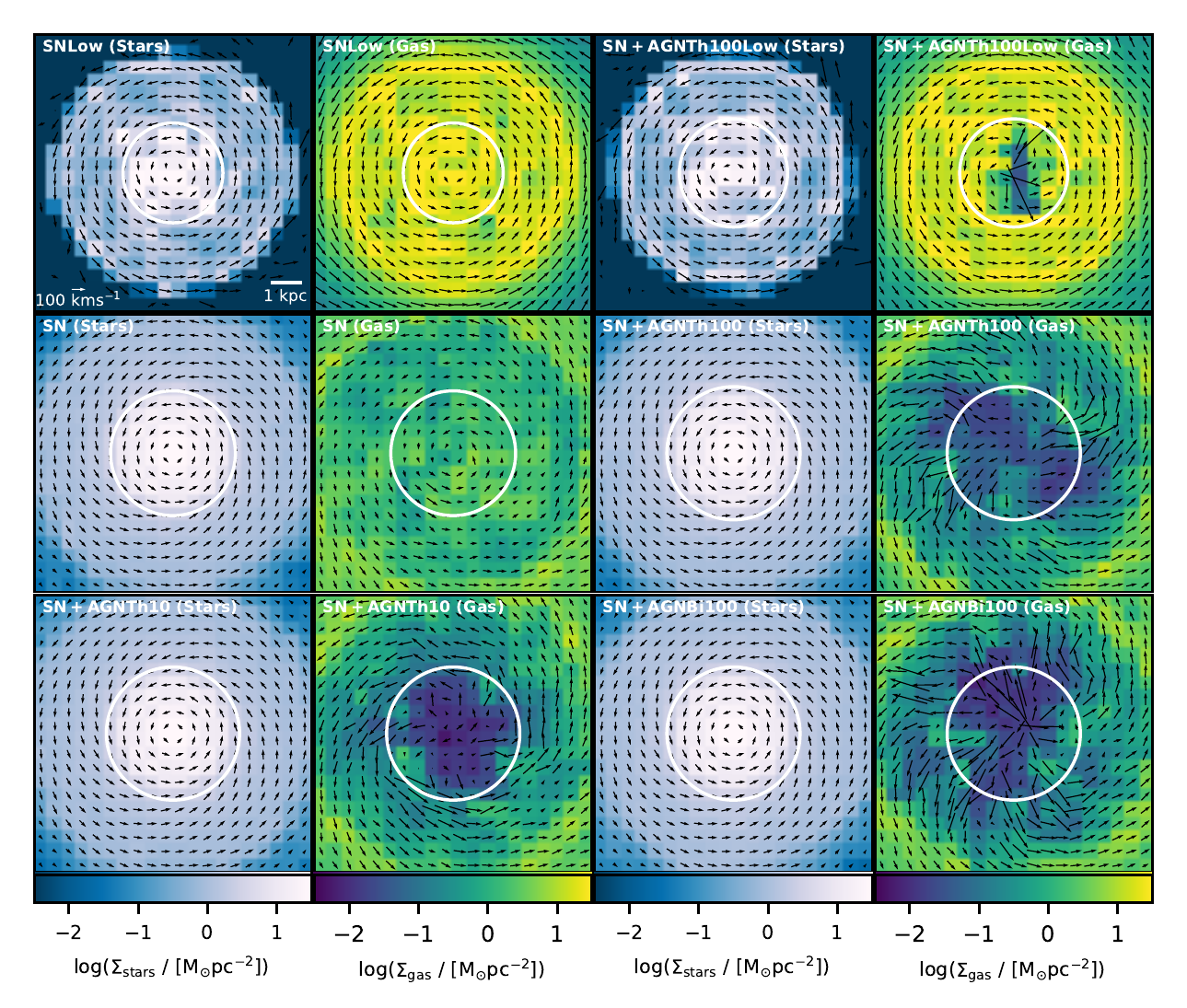}
\caption{Comparison between the 2D velocity maps of the stars and gas in the simulations at $t=150$ Myr. The colour coding of the pixels indicates the surface density. The spatial coverage of the primary MaNGA sample of $1.5R_\mathrm{eff}$ is indicated by a white circle. With added AGN feedback, the central circular motion is significantly altered by the outflows so that the gas would be seen as kinematically offset. With SN feedback alone, the circular motion remains dominant and we do not observe this feature.}
\label{fig:2D_velmaps} 
\end{figure*}
\begin{figure*}
\centering
\includegraphics[width=\textwidth]{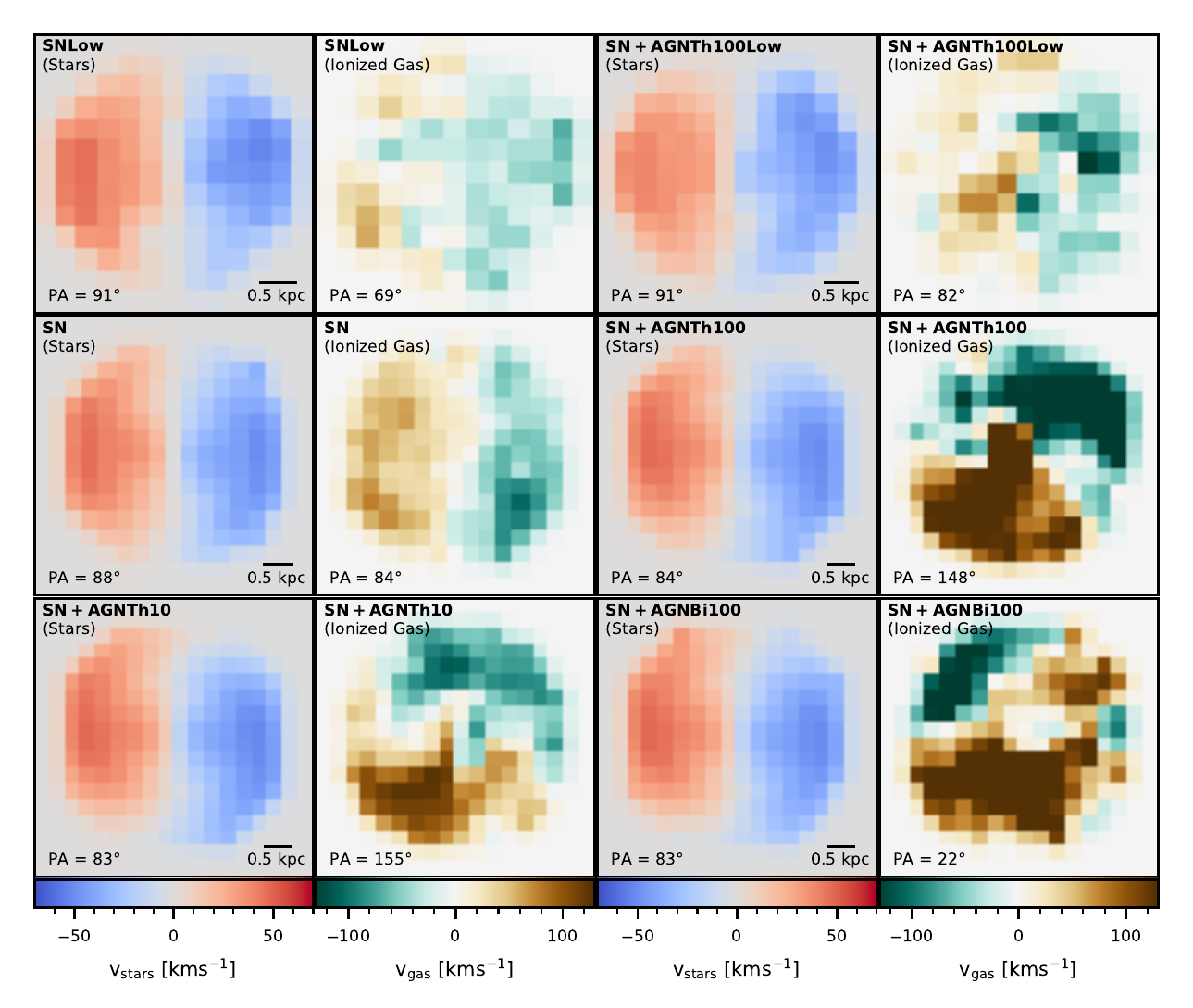}
\caption{Comparison between the line-of-sight velocity maps for stars and ionized gas at $t=150$ Myr. Only gas and stars within 1.5$\mathrm{R_\mathrm{eff}}$ are included (in analogy to MaNGA), and the spatial resolution is matched to the MaNGA resolution at the mean redshift of the primary sample. For the AGN runs, the ionized gas is visibly offset from the stellar component. }
\label{fig:los_velmaps} 
\end{figure*}
\begin{figure*}
\includegraphics[width=\textwidth, left]{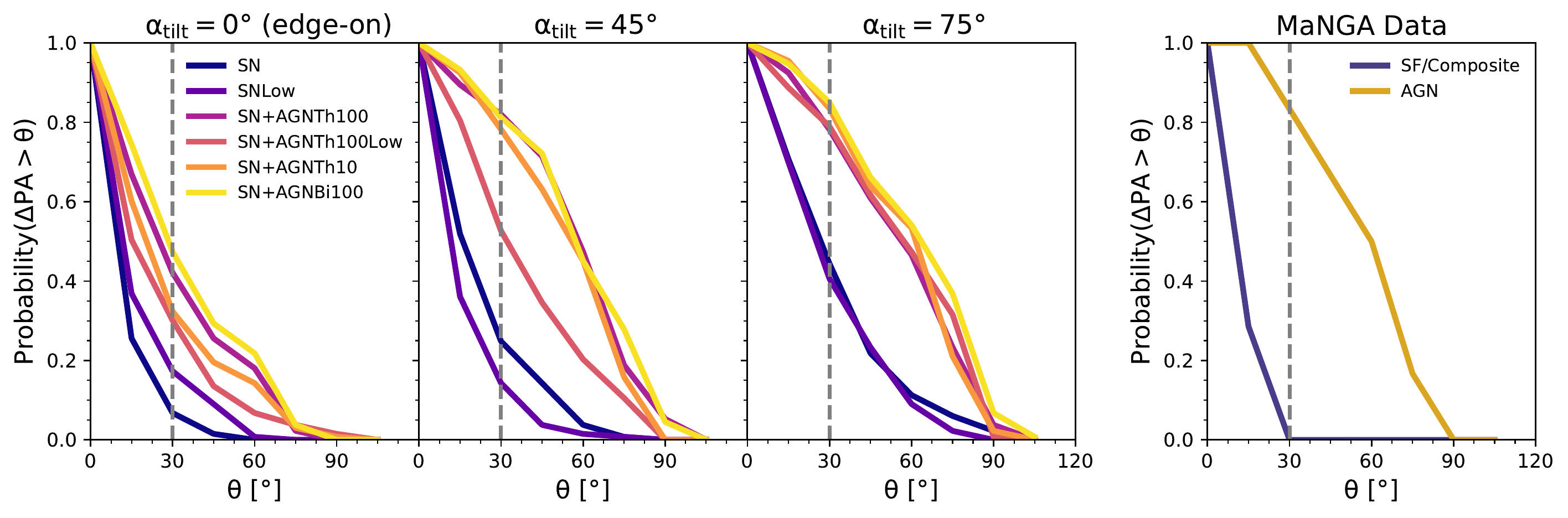}
\caption{Probability distribution for the kinematic offset between gas and stars to be at least an angle $\theta$ for different inclination angles of the disc from edge-on (0\degree) to nearly face-on (75\degree). These `probabilities' are calculated as normalised frequencies of how often a certain offset is observed for $t>40$ Myr. For the observed data, we use the thirteen quenched MaNGA galaxies from \citet{Penny2018a}, which also have a detectable ionized gas component. The minimum angle for the gas to be categorised as `kinematically offset' using the \citet{Penny2018a} scheme, $\theta=30\degree$, is shown as a grey dashed line.  For all three inclination angles galaxies with AGN activity are significantly more likely to be categorised as offset. }
\label{fig:delta_PA}
\end{figure*}
There is some first evidence of AGN feedback in dwarf galaxies from the SDSS-IV MaNGA survey. \citet{Penny2018a} analyse the MaNGA MPL-5 sample to find AGN candidates in the dwarf galaxy population\footnote{Here dwarf galaxies are defined as having absolute magnitude fainter than $M_\mathrm{V}= -19$, central stellar velocity dispersion $\sigma_{*}<100\ \mathrm{kms^{-1}}$ and stellar mass $M_{*}<5\times10^{9}\ \mathrm{M_{\odot}}$.}.  In particular they aim to find red geyser candidates, which typically have kinematically offset ionized gas, either the result of accretion or outflows of material generated by AGN-driven winds. They identify five dwarf galaxies with little or no star formation and an ionized gas component which is offset from the stars. All of these five galaxies have central spectral signatures indicating AGN activity, providing tentative evidence for maintenance mode AGN feedback.

We would like to test whether a similar kinematic offset would be observed for our simulated dwarf galaxies with AGN-driven outflows and how the probability of observing such an offset compares to the runs with only SNe. For an observable effect on the line-of-sight velocity, the outflow component has to be larger than the rotational component. In Figure \ref{fig:2D_velmaps}, we show some representative two-dimensional maps of the mass-weighted velocity components in the disc plane for gas and newly-formed stars (i.e. excluding the stars from the initial conditions) at $t=150$ Myr. The colour coding of the pixels indicates the surface density allowing us to relate the velocities to the matter distribution. In addition, the region covered by $1.5R_\mathrm{eff}$, where $R_\mathrm{eff}$ is the effective radius, corresponding to the spatial coverage of the primary MaNGA sample, is indicated by a white circle. For all runs, the rotational pattern of the stellar component is undisturbed, suggesting that the large-scale kinematics of the newly-formed stars are not affected by feedback or outflows. The gas velocities are not affected by SN feedback either. However, adding an AGN makes a significant difference, with the central velocity patterns being visibly altered by the AGN-driven outflows. As only the central region is taken into account when inferring line-of-sight velocities, the gas component will then be significantly offset from the stellar component.

To have a fair comparison between our simulation and the MaNGA results, we create mock line-of-sight velocity maps for both stars and ionized gas\footnote{Note that in our simulations ionized gas is the result of AGN heating while we do not consider the effect of AGN radiation and how this may lead to additional ionized gas.}. Here we define a gas cell as ionized if the neutral hydrogen abundance is less than 0.1 per cent. We have checked a range of values from $10^{-3}$ to one per cent and find that our results do not depend on this cut-off. When creating the velocity maps, we only use gas cells and stars within $1.5R_\mathrm{eff}$, in analogy to MaNGA. We also match the resolution of our velocity maps to the MaNGA resolution at the median redshift of the primary sample ($\sim304$ pc). In each pixel, we weight the velocities by the luminosity, i.e. by the mass for the stars and by the square of the density for the gas, as the latter is proportional to the emissivity. As the galaxies in the MaNGA sample have no preference for a particular orientation, we create velocity maps at three different inclination angles $\alpha_\mathrm{tilt}$: 0\degree (edge-on), 45\degree, and 75\degree (nearly face-on). We do not create velocity maps for the face-on case as the contribution of the rotational velocity to the line-of-sight velocity virtually vanishes.

We then carry out the same analysis that was done with the MaNGA maps in \citet{Penny2018a}. Firstly, we convolve the velocity maps with a Gaussian filter of filter size two pixels. We then use these maps to measure the global kinematic position angle\footnote{The kinematic position angle describes the orientation of the velocity map, and is related to the projection of the angular momentum vector.} (PA) of the ionized gas and the stellar component using the method described in \citet{Krajnovic2006}. Figure \ref{fig:los_velmaps} shows example mock velocity maps for $\alpha_\mathrm{tilt}=45\degree$ at $t=150$ Myr. The PAs are also given. As we have chosen the axes so that the angular momentum vector of the disc is aligned with the z axis, we would expect the PA to be 90\degree. For the stellar component this gets recovered within $\pm 10\degree$. Yet for the gas component, we obtain significant offsets when AGN activity is included, as could have been expected from Figure \ref{fig:2D_velmaps}. Also note that these non-rotational velocity components have much higher magnitudes than the rotational velocities. This effect is even more significant than in the observed MaNGA maps. However, the low-mass AGN host galaxies in MaNGA have a typical bolometric luminosity of $L_\mathrm{bol}=3.7\times 10^{41}\ \mathrm{ergs^{-1}}$ so with a BH mass of $10^5\ \mathrm{M_{\odot}}$, this would be approximately three per cent of the Eddington luminosity rather than the 10 or 100 per cent that we employ in our simulations.

We then go on to construct velocity maps and measure the kinematic offsets for simulation outputs every 2 Myr between $t=40$ Myr and $t=300$ Myr (we choose 40 Myr as a starting time to allow the initial conditions to settle and the SN feedback to develop fully). We then use all of these values to calculate the probability for the offset to be at least a given angle $\theta$ for the three different inclination angles. These probability distributions are shown in the first three panels of Figure \ref{fig:delta_PA}. The minimum angle for the gas to be categorised as `kinematically offset' using the \citet{Penny2018a} scheme, $\theta=30\degree$, is shown as a grey dashed line. For all three inclination angles, galaxies with AGN are significantly more likely to be categorised as offset, and this difference increases with increasing inclination angle.

In the rightmost panel we show the probability distribution inferred from the thirteen quenched MaNGA galaxies which have both a detectable ionized gas and stellar component\footnote{We adjust the kinematic offsets given by \citet{Penny2018a} to be within $0\degree<\Delta\mathrm{PA}<90\degree$ to facilitate comparison with our simulations which are always co-rotating.}. We separate out the seven galaxies which have been identified as star-forming (SF) or composite and the six galaxies which have been identified as AGN in the BPT diagram. As with our simulated galaxies, the probability to obtain significant offsets is much higher, and the frequency for different angles for the galaxies with an AGN is very similar to the distribution in our simulations. Though these probabilities should be treated with caution due to the small sample size. Also note that for MaNGA there is not a single galaxy without AGN activity which has been categorised as offset, while from our models we would expect between 10 and 30 per cent of SF galaxies to fall into that category. However, note that \citet{Penny2018a} only select galaxies which have little or no star formation so by construction there should not be strong stellar feedback present in this sample. If we impose an arbitrary SFR cut on our simulated data points, e.g. $\mathrm{SFR}<0.5\ \mathrm{M_{\odot} yr^{-1}}$, the probability to obtain significant kinematic offsets with the SN only runs is significantly lower, while the probability distribution of the AGN runs is barely affected. 

\section{Discussion} \label{section:Discussion}
We considered AGN-driven outflows in tandem with SN feedback to assess whether AGN activity affects dwarf galaxy properties. For this investigation, we made use of novel feedback models and a super-Lagrangian refinement scheme which allowed us to increase resolving power by up to three orders of magnitude in mass resolution around the black hole. We found that AGN feedback on its own only has a small effect on global SFRs. This is in line with findings from other numerical simulations, e.g. \citet{Schaye2015, Sijacki2007, Wadepuhl2010} find that including AGN feedback in addition to SN feedback only has a negligible effect on SFRs for low-mass systems. The analytical findings from \citet{Dashyan2018}, however, would suggest that AGN feedback should dominate over SN feedback for our set-up. While we do not obtain this, we find that AGN feedback has a significant effect on local quantities, e.g. the location of star formation and the driving of outflows. It seems that different from the distributed SN feedback, feedback from AGN is mainly local. The MaNGA survey \citep{Penny2018a} finds that dwarfs with observable signs of AGN activity have an ionized gas component that is kinematically offset from the stellar component suggesting that the AGN is causing outflows. We observe a similar effect in our simulations which further supports the supposition that this kinematic signature is AGN-driven.

However, note that we may be underestimating the effectiveness of SNe as we neither model other stellar feedback channels nor have a turbulent ISM model, which \citet{Hopkins2014} and \citet{Kimm2015} find to significantly enhance the ability of SNe to efficiently couple to the gas. This may reduce the gas supply to the central black hole or lead to hotter and faster outflows even with lower AGN luminosities. Further simulations are needed to explore these issues in detail.

Another shortcoming of our model is the isolated set-up meaning that we do not have gas inflows. The next step in this investigation would then be to run cosmological zoom-in simulations of AGN feedback in dwarf galaxies and investigate how the effectiveness of AGN feedback changes over cosmic time. \citet{Smith2018b} find that in their SN only simulations of high-redshift dwarf galaxies, significant amounts of dense gas accumulate in the galaxy centre. This high-redshift behaviour is worth exploring as high central gas densities would facilitate the growth of IMBHs in the early Universe and also make it easier for AGN feedback to regulate the overall SFR. Furthermore, the more irregular and clumpy morphology of high-redshift dwarfs may facilitate AGN outflow leakage into the CGM. Preventive feedback in the form of CGM heating by the hot AGN-driven outflows might play a more important role than ejective feedback for AGN in dwarf galaxies \citep[cf.][]{Penny2018a}.

Other studies of AGN feedback in cosmological simulations of dwarf galaxies have found that SN feedback stunts accretion \citep[see e.g.][]{Trebitsch2017,Habouzit2017}. However, we know that there are at least some AGN in dwarfs accreting efficiently from observations. To replicate this in numerical simulations we need a more elaborate accretion scheme than the currently widely adopted Bondi accretion model. 

\vspace{-4ex}
\section{Conclusion}\label{section:Conclusion}
We have investigated the role of AGN-driven outflows in dwarf galaxies using isolated disc galaxy simulations. The disc contains the majority of the baryonic mass with $M_\mathrm{disc}=3.5\times10^{9}\ \mathrm{M_{\odot}}$, and is gas-rich: 50 per cent of the initial disc mass is in gas akin to high-redshift analogues. We add a black hole of mass $M_\mathrm{BH}=10^5\ \mathrm{M_{\odot}}$ allowing us to test AGN activity at the upper end of the IMBH range.

For this investigation, we made use of the star formation and SN feedback model by \citet{Smith2018} and the AGN model by \citet{Curtis2015}. We used different AGN outflow models, ranging from simple energy-driven spherical winds to collimated, mass-loaded, bipolar outflows, and employed a super-Lagrangian refinement scheme to accurately resolve the energy injection, increasing the gas resolution around the central black hole by three orders of magnitude. We tested two different star formation efficiencies, $\epsilon_\mathrm{SF}=0.015,0.15$ and two different (constant) AGN luminosities $L_\mathrm{AGN}=0.1L_\mathrm{Edd},1.0L_\mathrm{Edd}$. This way we could assess the maximum possible impact an AGN could have on this type of system. Our most important findings are the following:
\begin{enumerate}[(i)]
  \item There is a small but systematic effect of AGN outflows on central SFRs for all set-ups explored, while significant effects on the global SFR are only obtained with strong SN feedback and sustained high-luminosity isotropic AGN outflows (`SN+AGNTh100' run). This indicates that AGN feedback in dwarf galaxies is unlikely to directly regulate their global SFRs.
  \item AGN activity in dwarf galaxies is however effective in enhancing outflows when there is also strong SN feedback present. If the SN feedback is inefficient, even adding an AGN shining at 100 per cent of the Eddington luminosity does not produce large mass loading factors.
    \item With AGN feedback included, the outflows reach much higher velocities (up to $10^4\ \mathrm{kms^{-1}}$ compared to $10^2\ \mathrm{kms^{-1}}$ without an AGN) and much higher temperatures (up to $10^9$ K instead of $10^7$ K).
  \item We find similar kinematic signatures as in observed dwarf galaxies hosting an AGN from the MaNGA survey. SN feedback alone does not drive strong enough outflows to significantly offset the rotational motion - even when the star formation efficiency is high and SN feedback is strong. This then strengthens the supposition by \citet{Penny2018a} that these kinematic offsets are caused by an AGN-driven outflow.
\end{enumerate}
Cosmological simulations are needed to fully answer the question of whether AGN feedback could affect star formation in dwarf galaxies. Our findings from the isolated galaxy set-up indicate that while AGN are unlikely to directly affect global dwarf SFRs, galactic outflows are significantly enhanced. In realistic cosmological environments inflows are known to be important especially for high-redshift dwarfs. It is hence possible that AGN-boosted outflows may prevent some of this cosmic `pristine' gas reaching the dwarfs in the first place, providing a mechanism for indirect star formation regulation. 
\vspace{-4ex}
\section{Acknowledgements}
We are grateful to Martin Haehnelt and Renske Smit for helpful comments. We also thank Roberto Maiolino and Robert Gallagher for useful discussions on kinematic studies with MaNGA. SK, DS, MAB and MCS acknowledge support by the Science and Technology Facilities Council (STFC) and the ERC Starting Grant 638707 ``Black holes and their host galaxies: co-evolution across cosmic time''. This work was performed on the following: the Cambridge Service for Data Driven Discovery (CSD3), part of which is operated by the University of Cambridge Research Computing on behalf of the STFC DiRAC HPC Facility (www.dirac.ac.uk). The DiRAC component of CSD3 was funded by BEIS capital funding via STFC capital grants ST/P002307/1 and ST/R002452/1 and STFC operations grant ST/R00689X/1; the DiRAC@Durham facility managed by the Institute for Computational Cosmology on behalf of the STFC DiRAC HPC Facility (www.dirac.ac.uk). The equipment was funded by BEIS capital funding via STFC capital grants ST/P002293/1 and ST/R002371/1, Durham University and STFC operations grant ST/R000832/1. DiRAC is part of the National e-Infrastructure.
\vspace{-4ex}
\bibliographystyle{mn2e} 
\bibliography{./sk_2018.bib}

\begin{thebibliography}{}
\makeatletter
\relax
\def\mn@urlcharsother{\let\do\@makeother \do\$\do\&\do\#\do\^\do\_\do\%\do\~}
\def\mn@doi{\begingroup\mn@urlcharsother \@ifnextchar [ {\mn@doi@}
  {\mn@doi@[]}}
\def\mn@doi@[#1]#2{\def\@tempa{#1}\ifx\@tempa\@empty \href
  {http://dx.doi.org/#2} {doi:#2}\else \href {http://dx.doi.org/#2} {#1}\fi
  \endgroup}
\def\mn@eprint#1#2{\mn@eprint@#1:#2::\@nil}
\def\mn@eprint@arXiv#1{\href {http://arxiv.org/abs/#1} {{\tt arXiv:#1}}}
\def\mn@eprint@dblp#1{\href {http://dblp.uni-trier.de/rec/bibtex/#1.xml}
  {dblp:#1}}
\def\mn@eprint@#1:#2:#3:#4\@nil{\def\@tempa {#1}\def\@tempb {#2}\def\@tempc
  {#3}\ifx \@tempc \@empty \let \@tempc \@tempb \let \@tempb \@tempa \fi \ifx
  \@tempb \@empty \def\@tempb {arXiv}\fi \@ifundefined
  {mn@eprint@\@tempb}{\@tempb:\@tempc}{\expandafter \expandafter \csname
  mn@eprint@\@tempb\endcsname \expandafter{\@tempc}}}

\bibitem[\protect\citeauthoryear{Baldassare et~al.,}{Baldassare
  et~al.}{2016}]{Baldassare2016}
Baldassare V.~F.,  et~al., 2016, \mn@doi [ApJ] {10.3847/0004-637X/829/1/57},
  829, 57

\bibitem[\protect\citeauthoryear{Baldassare, Reines, Gallo  \&
  Greene}{Baldassare et~al.}{2017}]{Baldassare2016a}
Baldassare V.~F.,  Reines A.~E.,  Gallo E.,   Greene J.~E.,  2017, \mn@doi
  [ApJ] {10.3847/1538-4357/836/1/20}, 836, 20

\bibitem[\protect\citeauthoryear{Baldwin, Phillips  \& Terlevich}{Baldwin
  et~al.}{1981}]{Baldwin1981}
Baldwin A.,  Phillips M.~M.,   Terlevich R.,  1981, \mn@doi [PASP]
  {10.1086/130930}, 93, 817

\bibitem[\protect\citeauthoryear{Barai \& Pino}{Barai \&
  Pino}{2018}]{Barai2018a}
Barai P.,  Pino E. M. d. G.~D.,  2018, preprint (\mn@eprint {arXiv}
  {1807.04768})

\bibitem[\protect\citeauthoryear{Beckmann et~al.,}{Beckmann
  et~al.}{2017}]{Beckmann2017}
Beckmann R.~S.,  et~al., 2017, \mn@doi [MNRAS] {10.1093/mnras/stx1831}, 472,
  949

\bibitem[\protect\citeauthoryear{Bellovary, Cleary, Munshi, Tremmel,
  Christensen, Brooks  \& Quinn}{Bellovary et~al.}{2018}]{Bellovary2018}
Bellovary J.,  Cleary C.,  Munshi F.,  Tremmel M.,  Christensen C.,  Brooks A.,
    Quinn T.,  2018, \mn@doi [MNRAS] {10.1093/mnras/sty2842}, 482, 2913

\bibitem[\protect\citeauthoryear{Bieri, Dubois, Rosdahl, Wagner, Silk  \&
  Mamon}{Bieri et~al.}{2017}]{Bieri2017}
Bieri R.,  Dubois Y.,  Rosdahl J.,  Wagner A.,  Silk J.,   Mamon G.~A.,  2017,
  \mn@doi [MNRAS] {10.1093/mnras/stw2380}, 464, 1854

\bibitem[\protect\citeauthoryear{Binney \& Tabor}{Binney \&
  Tabor}{1995}]{Binney1995}
Binney J.,  Tabor G.,  1995, \mn@doi [MNRAS] {10.1093/mnras/276.2.663}, 276,
  663

\bibitem[\protect\citeauthoryear{Blanton et~al.,}{Blanton
  et~al.}{2017}]{Blanton2017}
Blanton M.~R.,  et~al., 2017, \mn@doi [AJ] {10.3847/1538-3881/aa7567}, 154, 28

\bibitem[\protect\citeauthoryear{Blondin, Wright, Borkowski  \&
  Reynolds}{Blondin et~al.}{1998}]{Blondin1998}
Blondin J.~M.,  Wright E.~B.,  Borkowski K.~J.,   Reynolds S.~P.,  1998,
  \mn@doi [ApJ] {10.1086/305708}, 500, 342

\bibitem[\protect\citeauthoryear{Bondi \& Hoyle}{Bondi \&
  Hoyle}{1944}]{Bondi1944}
Bondi H.,  Hoyle F.,  1944, \mn@doi [MNRAS] {10.1093/mnras/104.5.273}, 104, 273

\bibitem[\protect\citeauthoryear{Bourne, Nayakshin  \& Hobbs}{Bourne
  et~al.}{2014}]{Bourne2014}
Bourne M.~A.,  Nayakshin S.,   Hobbs A.,  2014, \mn@doi [MNRAS]
  {10.1093/mnras/stu747}, 441, 3055

\bibitem[\protect\citeauthoryear{Bower, Benson, Malbon, Helly, Frenk, Baugh,
  Cole  \& Lacey}{Bower et~al.}{2006}]{Bower2006}
Bower R.~G.,  Benson A.~J.,  Malbon R.,  Helly J.~C.,  Frenk C.~S.,  Baugh
  C.~M.,  Cole S.,   Lacey C.~G.,  2006, \mn@doi [MNRAS]
  {10.1111/j.1365-2966.2006.10519.x}, 370, 645

\bibitem[\protect\citeauthoryear{Boylan-Kolchin, Bullock  \&
  Kaplinghat}{Boylan-Kolchin et~al.}{2011}]{Boylan-Kolchin2011}
Boylan-Kolchin M.,  Bullock J.~S.,   Kaplinghat M.,  2011, \mn@doi [MNRAS]
  {10.1111/j.1745-3933.2011.01074.x}, 415, L40

\bibitem[\protect\citeauthoryear{Bundy et~al.,}{Bundy et~al.}{2014}]{Bundy2015}
Bundy K.,  et~al., 2014, \mn@doi [ApJ] {10.1088/0004-637X/798/1/7}, 798, 7

\bibitem[\protect\citeauthoryear{Chilingarian, Katkov, Zolotukhin, Grishin,
  Beletsky, Boutsia  \& Osip}{Chilingarian et~al.}{2018}]{Chilingarian2018a}
Chilingarian I.~V.,  Katkov I.~Y.,  Zolotukhin I.~Y.,  Grishin K.~A.,  Beletsky
  Y.,  Boutsia K.,   Osip D.~J.,  2018, \mn@doi [ApJ] {arXiv:1805.01467v1},
  863, 1

\bibitem[\protect\citeauthoryear{Choi, Ostriker, Naab  \& Johansson}{Choi
  et~al.}{2012}]{Choi2012}
Choi E.,  Ostriker J.~P.,  Naab T.,   Johansson P.~H.,  2012, \mn@doi [ApJ]
  {10.1088/0004-637X/754/2/125}, 754, 125

\bibitem[\protect\citeauthoryear{Choi, Ostriker, Naab, Oser  \& Moster}{Choi
  et~al.}{2015}]{Choi2015}
Choi E.,  Ostriker J.~P.,  Naab T.,  Oser L.,   Moster B.~P.,  2015, \mn@doi
  [MNRAS] {10.1093/mnras/stv575}, 449, 4105

\bibitem[\protect\citeauthoryear{Ciardi, Salvaterra  \& {Di Matteo}}{Ciardi
  et~al.}{2010}]{Ciardi2010}
Ciardi B.,  Salvaterra R.,   {Di Matteo} T.,  2010, \mn@doi [MNRAS]
  {10.1111/j.1365-2966.2009.15843.x}, 401, 2635

\bibitem[\protect\citeauthoryear{Costa, Sijacki  \& Haehnelt}{Costa
  et~al.}{2014}]{Costa2014}
Costa T.,  Sijacki D.,   Haehnelt M.~G.,  2014, \mn@doi [MNRAS]
  {10.1093/mnras/stu1632}, 444, 2355

\bibitem[\protect\citeauthoryear{Costa, Rosdahl, Sijacki  \& Haehnelt}{Costa
  et~al.}{2018}]{Costa2018}
Costa T.,  Rosdahl J.,  Sijacki D.,   Haehnelt M.~G.,  2018, \mn@doi [MNRAS]
  {10.1093/mnras/stx2598}, 473, 4197

\bibitem[\protect\citeauthoryear{Croton}{Croton}{2006}]{Croton2006}
Croton D.~J.,  2006, \mn@doi [MNRAS] {10.1111/j.1365-2966.2006.10429.x}, 369,
  1808

\bibitem[\protect\citeauthoryear{Curtis \& Sijacki}{Curtis \&
  Sijacki}{2015}]{Curtis2015}
Curtis M.,  Sijacki D.,  2015, \mn@doi [MNRAS] {10.1093/mnras/stv2246}, 454,
  3445

\bibitem[\protect\citeauthoryear{Dashyan, Silk, Mamon, Dubois  \&
  Hartwig}{Dashyan et~al.}{2018}]{Dashyan2018}
Dashyan G.,  Silk J.,  Mamon G.~A.,  Dubois Y.,   Hartwig T.,  2018, \mn@doi
  [MNRAS] {10.1093/mnras/stx2716}, 473, 5698

\bibitem[\protect\citeauthoryear{Debuhr, Quataert  \& Ma}{Debuhr
  et~al.}{2011}]{DeBuhr2010}
Debuhr J.,  Quataert E.,   Ma C.~P.,  2011, \mn@doi [MNRAS]
  {10.1111/j.1365-2966.2010.17992.x}, 412, 1341

\bibitem[\protect\citeauthoryear{Dekel \& Silk}{Dekel \&
  Silk}{1986}]{Dekel1986}
Dekel A.,  Silk J.,  1986, \mn@doi [ApJ] {10.1086/164050}, 303, 39

\bibitem[\protect\citeauthoryear{Desroches, Greene  \& Ho}{Desroches
  et~al.}{2009}]{Desroches2009}
Desroches L.-B.,  Greene J.~E.,   Ho L.~C.,  2009, \mn@doi [ApJ]
  {10.1088/0004-637X/698/2/1515}, 698, 1515

\bibitem[\protect\citeauthoryear{{Di Matteo}, Springel  \& Hernquist}{{Di
  Matteo} et~al.}{2005}]{DiMatteo2005}
{Di Matteo} T.,  Springel V.,   Hernquist L.,  2005, \mn@doi [Nature]
  {10.1038/nature03335}, 433, 604

\bibitem[\protect\citeauthoryear{Dong, Greene  \& Ho}{Dong
  et~al.}{2012}]{Dong2012}
Dong R.,  Greene J.~E.,   Ho L.~C.,  2012, \mn@doi [ApJ]
  {10.1088/0004-637X/761/1/73}, 761, 73

\bibitem[\protect\citeauthoryear{Efstathiou}{Efstathiou}{1992}]{Efstathiou1992}
Efstathiou G.,  1992, \mn@doi [MNRAS] {10.1093/mnras/256.1.43P}, 256, 43P

\bibitem[\protect\citeauthoryear{Emerick, Bryan  \& {Mac Low}}{Emerick
  et~al.}{2018}]{Emerick2018}
Emerick A.,  Bryan G.~L.,   {Mac Low} M.-M.,  2018, \mn@doi [ApJ]
  {10.3847/2041-8213/aae315}, 865, L22

\bibitem[\protect\citeauthoryear{Fitts et~al.,}{Fitts et~al.}{2016}]{Fitts2017}
Fitts A.,  et~al., 2016, \mn@doi [MNRAS] {10.1093/mnras/stx1757}, 471, 3547

\bibitem[\protect\citeauthoryear{Gabor \& Bournaud}{Gabor \&
  Bournaud}{2014}]{Gabor2014}
Gabor J.~M.,  Bournaud F.,  2014, \mn@doi [MNRAS] {10.1093/mnras/stu677}, 441,
  1615

\bibitem[\protect\citeauthoryear{Garrison-Kimmel, Boylan-Kolchin, Bullock  \&
  Kirby}{Garrison-Kimmel et~al.}{2014}]{Garrison-Kimmel2014a}
Garrison-Kimmel S.,  Boylan-Kolchin M.,  Bullock J.~S.,   Kirby E.~N.,  2014,
  \mn@doi [MNRAS] {10.1177/0958305X17750052}, 444, 222

\bibitem[\protect\citeauthoryear{Geen, Rosdahl, Blaizot, Devriendt  \&
  Slyz}{Geen et~al.}{2015}]{Geen2015}
Geen S.,  Rosdahl J.,  Blaizot J.,  Devriendt J.,   Slyz A.,  2015, \mn@doi
  [MNRAS] {10.1093/mnras/stv251}, 448, 3248

\bibitem[\protect\citeauthoryear{Governato et~al.,}{Governato
  et~al.}{2010}]{Governato2010}
Governato F.,  et~al., 2010, \mn@doi [Nature] {10.1038/nature08640}, 463, 203

\bibitem[\protect\citeauthoryear{Graham \& Soria}{Graham \&
  Soria}{2018}]{Graham2018b}
Graham A.~W.,  Soria R.,  2018, preprint (\mn@eprint {arXiv} {1812.01231})

\bibitem[\protect\citeauthoryear{Graham, Soria  \& Davis}{Graham
  et~al.}{2018}]{Graham2018a}
Graham A.~W.,  Soria R.,   Davis B.~L.,  2018, preprint (\mn@eprint {arXiv}
  {1811.03232})

\bibitem[\protect\citeauthoryear{Greene \& Ho}{Greene \& Ho}{2004}]{Greene2004}
Greene J.~E.,  Ho L.~C.,  2004, \mn@doi [ApJ] {10.1086/421719}, 610, 13

\bibitem[\protect\citeauthoryear{Greene \& Ho}{Greene \& Ho}{2007}]{Greene2007}
Greene J.~E.,  Ho L.~C.,  2007, \mn@doi [ApJ] {10.1086/522082}, 670, 92

\bibitem[\protect\citeauthoryear{Habouzit, Volonteri  \& Dubois}{Habouzit
  et~al.}{2017}]{Habouzit2017}
Habouzit M.,  Volonteri M.,   Dubois Y.,  2017, \mn@doi [MNRAS]
  {10.1093/mnras/stx666}, 468, 3935

\bibitem[\protect\citeauthoryear{Henden, Puchwein, Shen  \& Sijacki}{Henden
  et~al.}{2018}]{Henden2018}
Henden N.~A.,  Puchwein E.,  Shen S.,   Sijacki D.,  2018, \mn@doi [MNRAS]
  {10.1093/mnras/sty1780}, 479, 5385

\bibitem[\protect\citeauthoryear{Hopkins, Richards  \& Hernquist}{Hopkins
  et~al.}{2007}]{Hopkins2006}
Hopkins P.~F.,  Richards G.~T.,   Hernquist L.,  2007, \mn@doi [ApJ]
  {10.1086/509629}, 654, 731

\bibitem[\protect\citeauthoryear{Hopkins, Kere{\v{s}}, O{\~{n}}orbe,
  Faucher-Gigu{\`{e}}re, Quataert, Murray  \& Bullock}{Hopkins
  et~al.}{2014}]{Hopkins2014}
Hopkins P.~F.,  Kere{\v{s}} D.,  O{\~{n}}orbe J.,  Faucher-Gigu{\`{e}}re C.~A.,
   Quataert E.,  Murray N.,   Bullock J.~S.,  2014, \mn@doi [MNRAS]
  {10.1093/mnras/stu1738}, 445, 581

\bibitem[\protect\citeauthoryear{Hopkins et~al.,}{Hopkins
  et~al.}{2018a}]{Hopkins2018}
Hopkins P.~F.,  et~al., 2018a, \mn@doi [MNRAS] {10.1093/mnras/sty674}, 477,
  1578

\bibitem[\protect\citeauthoryear{Hopkins et~al.,}{Hopkins
  et~al.}{2018b}]{Hopkins2018a}
Hopkins P.~F.,  et~al., 2018b, \mn@doi [MNRAS] {10.1093/mnras/sty1690}, 480,
  800

\bibitem[\protect\citeauthoryear{Hoyle \& Lyttleton}{Hoyle \&
  Lyttleton}{1939}]{Hoyle1939}
Hoyle F.,  Lyttleton R.~A.,  1939, \mn@doi [Math. Proc. Cambridge Philos. Soc.]
  {10.1017/S0305004100021150}, 35, 405

\bibitem[\protect\citeauthoryear{Hu}{Hu}{2018}]{Hu2018}
Hu C.-Y.,  2018, \mn@doi [MNRAS] {10.1093/mnras/sty3252}, 264, 21923

\bibitem[\protect\citeauthoryear{Kauffmann, White  \& Guiderdoni}{Kauffmann
  et~al.}{1993}]{Kauffmann1993}
Kauffmann G.,  White S. D.~M.,   Guiderdoni B.,  1993, \mn@doi [MNRAS]
  {10.1093/mnras/264.1.201}, 264, 201

\bibitem[\protect\citeauthoryear{Kim \& Ostriker}{Kim \&
  Ostriker}{2015}]{Kim2015}
Kim C.~G.,  Ostriker E.~C.,  2015, \mn@doi [ApJ] {10.1088/0004-637X/802/2/99},
  802, 99

\bibitem[\protect\citeauthoryear{Kimm, Cen, Devriendt, Dubois  \& Slyz}{Kimm
  et~al.}{2015}]{Kimm2015}
Kimm T.,  Cen R.,  Devriendt J.,  Dubois Y.,   Slyz A.,  2015, \mn@doi [MNRAS]
  {10.1093/mnras/stv1211}, 451, 2900

\bibitem[\protect\citeauthoryear{Klypin, Kravtsov, Valenzuela  \& Prada}{Klypin
  et~al.}{1999}]{Klypin1999}
Klypin A.~A.,  Kravtsov A.~V.,  Valenzuela O.,   Prada F.,  1999, \mn@doi [ApJ]
  {10.1086/307643}, 522, 82

\bibitem[\protect\citeauthoryear{Kormendy \& Ho}{Kormendy \&
  Ho}{2013}]{Kormendy2013}
Kormendy J.,  Ho L.~C.,  2013, \mn@doi [ARA{\&}A]
  {10.1146/annurev-astro-082708-101811}, 51, 511

\bibitem[\protect\citeauthoryear{Krajnovi{\'{c}}, Cappellari, {De Zeeuw}  \&
  Copin}{Krajnovi{\'{c}} et~al.}{2006}]{Krajnovic2006}
Krajnovi{\'{c}} D.,  Cappellari M.,  {De Zeeuw} P.~T.,   Copin Y.,  2006,
  \mn@doi [MNRAS] {10.1111/j.1365-2966.2005.09902.x}, 366, 787

\bibitem[\protect\citeauthoryear{Kroupa}{Kroupa}{2002}]{Kroupa2002}
Kroupa P.,  2002, \mn@doi [Science] {10.1126/science.1067524}, 295, 82

\bibitem[\protect\citeauthoryear{Krumholz \& Tan}{Krumholz \&
  Tan}{2007}]{Krumholz2007}
Krumholz M.~R.,  Tan J.~C.,  2007, \mn@doi [ApJ] {10.1086/509101}, 654, 304

\bibitem[\protect\citeauthoryear{Leitherer et~al.,}{Leitherer
  et~al.}{1999}]{Leitherer1999}
Leitherer C.,  et~al., 1999, \mn@doi [ApJS] {10.1086/313233}, 123, 3

\bibitem[\protect\citeauthoryear{Lemons, Reines, Plotkin, Gallo  \&
  Greene}{Lemons et~al.}{2015}]{Lemons2015}
Lemons S.~M.,  Reines A.~E.,  Plotkin R.~M.,  Gallo E.,   Greene J.~E.,  2015,
  \mn@doi [ApJ] {10.1088/0004-637X/805/1/12}, 805, 12

\bibitem[\protect\citeauthoryear{Lovell et~al.,}{Lovell
  et~al.}{2012}]{Lovell2012}
Lovell M.~R.,  et~al., 2012, \mn@doi [MNRAS]
  {10.1111/j.1365-2966.2011.20200.x}, 420, 2318

\bibitem[\protect\citeauthoryear{Maiolino et~al.,}{Maiolino
  et~al.}{2012}]{Maiolino2012}
Maiolino R.,  et~al., 2012, \mn@doi [MNRAS] {10.1111/j.1745-3933.2012.01303.x},
  425, L66

\bibitem[\protect\citeauthoryear{Marleau, Clancy  \& Bianconi}{Marleau
  et~al.}{2013}]{Marleau2013}
Marleau F.~R.,  Clancy D.,   Bianconi M.,  2013, \mn@doi [MNRAS]
  {10.1093/mnras/stt1503}, 435, 3085

\bibitem[\protect\citeauthoryear{Martizzi, Faucher-Gigu{\`{e}}re  \&
  Quataert}{Martizzi et~al.}{2015}]{Martizzi2015}
Martizzi D.,  Faucher-Gigu{\`{e}}re C.-A.,   Quataert E.,  2015, \mn@doi
  [MNRAS] {10.1093/mnras/stv562}, 450, 504

\bibitem[\protect\citeauthoryear{Moore}{Moore}{1994}]{Moore1994}
Moore B.,  1994, \mn@doi [Nature] {10.1038/370629a0}, 370, 629

\bibitem[\protect\citeauthoryear{Moore, Ghigna, Governato, Lake, Quinn, Stadel
  \& Tozzi}{Moore et~al.}{1999}]{Moore1999}
Moore B.,  Ghigna S.,  Governato F.,  Lake G.,  Quinn T.,  Stadel J.,   Tozzi
  P.,  1999, \mn@doi [ApJ] {10.1086/312287}, 524, L19

\bibitem[\protect\citeauthoryear{Navarro, Eke  \& Frenk}{Navarro
  et~al.}{1996a}]{Navarro1996}
Navarro J.~F.,  Eke V.~R.,   Frenk C.~S.,  1996a, \mn@doi [MNRAS]
  {10.1093/mnras/283.3.L72}, 283, L72

\bibitem[\protect\citeauthoryear{Navarro, Frenk  \& White}{Navarro
  et~al.}{1996b}]{Navarro1997}
Navarro J.~F.,  Frenk C.~S.,   White S. D.~M.,  1996b, \mn@doi [ApJ]
  {10.1086/304888}, 490, 493

\bibitem[\protect\citeauthoryear{Nayakshin \& Zubovas}{Nayakshin \&
  Zubovas}{2012}]{Nayakshin2012}
Nayakshin S.,  Zubovas K.,  2012, \mn@doi [MNRAS]
  {10.1111/j.1365-2966.2012.21950.x}, 427, 372

\bibitem[\protect\citeauthoryear{Okamoto, Gao  \& Theuns}{Okamoto
  et~al.}{2008}]{Okamoto2008}
Okamoto T.,  Gao L.,   Theuns T.,  2008, \mn@doi [MNRAS]
  {10.1111/j.1365-2966.2008.13830.x}, 390, 920

\bibitem[\protect\citeauthoryear{Pardo et~al.,}{Pardo et~al.}{2016}]{Pardo2016}
Pardo K.,  et~al., 2016, \mn@doi [ApJ] {10.3847/0004-637X/831/2/203}, 831, 203

\bibitem[\protect\citeauthoryear{Parry, Eke, Frenk  \& Okamoto}{Parry
  et~al.}{2012}]{Parry2012}
Parry O.~H.,  Eke V.~R.,  Frenk C.~S.,   Okamoto T.,  2012, \mn@doi [MNRAS]
  {10.1111/j.1365-2966.2011.19971.x}, 419, 3304

\bibitem[\protect\citeauthoryear{Penny et~al.,}{Penny
  et~al.}{2018}]{Penny2018a}
Penny S.~J.,  et~al., 2018, \mn@doi [MNRAS] {10.1093/mnras/sty202}, 476, 979

\bibitem[\protect\citeauthoryear{Pontzen \& Governato}{Pontzen \&
  Governato}{2014}]{Pontzen2014}
Pontzen A.,  Governato F.,  2014, \mn@doi [Nature] {10.1038/nature12953}, 506,
  171

\bibitem[\protect\citeauthoryear{Puchwein \& Springel}{Puchwein \&
  Springel}{2012}]{Puchwein2013}
Puchwein E.,  Springel V.,  2012, \mn@doi [MNRAS] {10.1093/mnras/sts243}, 428,
  2966

\bibitem[\protect\citeauthoryear{Reines \& Volonteri}{Reines \&
  Volonteri}{2015}]{Reines2015}
Reines A.~E.,  Volonteri M.,  2015, \mn@doi [ApJ] {10.1088/0004-637X/813/2/82},
  813, 82

\bibitem[\protect\citeauthoryear{Reines, Greene  \& Geha}{Reines
  et~al.}{2013}]{Reines2013}
Reines A.~E.,  Greene J.~E.,   Geha M.,  2013, \mn@doi [ApJ]
  {10.1088/0004-637X/775/2/116}, 775

\bibitem[\protect\citeauthoryear{Rupke \& Veilleux}{Rupke \&
  Veilleux}{2011}]{Rupke2011}
Rupke D.~S.,  Veilleux S.,  2011, \mn@doi [ApJ] {10.1088/2041-8205/729/2/L27},
  729, 27

\bibitem[\protect\citeauthoryear{Sartori, Schawinski, Treister, Trakhtenbrot,
  Koss, Shirazi  \& Oh}{Sartori et~al.}{2015}]{Sartori2015}
Sartori L.~F.,  Schawinski K.,  Treister E.,  Trakhtenbrot B.,  Koss M.,
  Shirazi M.,   Oh K.,  2015, \mn@doi [MNRAS] {10.1093/mnras/stv2238}, 454,
  3722

\bibitem[\protect\citeauthoryear{Satyapal, Vega, Heckman, O'Halloran  \&
  Dudik}{Satyapal et~al.}{2007}]{Satyapal2007}
Satyapal S.,  Vega D.,  Heckman T.,  O'Halloran B.,   Dudik R.,  2007, \mn@doi
  [ApJ] {10.1086/519995}, 663, L9

\bibitem[\protect\citeauthoryear{Satyapal, Vega, Dudik, Abel  \&
  Heckman}{Satyapal et~al.}{2008}]{Satyapal2008}
Satyapal S.,  Vega D.,  Dudik R.~P.,  Abel N.~P.,   Heckman T.,  2008, \mn@doi
  [ApJ] {10.1086/529014}, 677, 17

\bibitem[\protect\citeauthoryear{Satyapal, Secrest, McAlpine, Ellison, Fischer
  \& Rosenberg}{Satyapal et~al.}{2014}]{Satyapal2014}
Satyapal S.,  Secrest N.~J.,  McAlpine W.,  Ellison S.~L.,  Fischer J.,
  Rosenberg J.~L.,  2014, \mn@doi [ApJ] {10.1088/0004-637X/784/2/113}, 784, 113

\bibitem[\protect\citeauthoryear{Sawala et~al.,}{Sawala
  et~al.}{2016}]{Sawala2016}
Sawala T.,  et~al., 2016, \mn@doi [MNRAS] {10.1093/mnras/stw145}, 457, 1931

\bibitem[\protect\citeauthoryear{Schaye et~al.,}{Schaye
  et~al.}{2015}]{Schaye2015}
Schaye J.,  et~al., 2015, \mn@doi [MNRAS] {10.1093/mnras/stu2058}, 446, 521

\bibitem[\protect\citeauthoryear{Sijacki, Springel, {Di Matteo}  \&
  Hernquist}{Sijacki et~al.}{2007}]{Sijacki2007}
Sijacki D.,  Springel V.,  {Di Matteo} T.,   Hernquist L.,  2007, \mn@doi
  [MNRAS] {10.1111/j.1365-2966.2007.12153.x}, 380, 877

\bibitem[\protect\citeauthoryear{Silk}{Silk}{2017}]{Silk2017}
Silk J.,  2017, \mn@doi [ApJ] {10.3847/2041-8213/aa67da}, 839, 13

\bibitem[\protect\citeauthoryear{Smith, Sijacki  \& Shen}{Smith
  et~al.}{2018a}]{Smith2018}
Smith M.~C.,  Sijacki D.,   Shen S.,  2018a, \mn@doi [MNRAS]
  {10.1093/mnras/sty994}, 478, 302

\bibitem[\protect\citeauthoryear{Smith, Sijacki  \& Shen}{Smith
  et~al.}{2018b}]{Smith2018b}
Smith M.~C.,  Sijacki D.,   Shen S.,  2018b, preprint, \GG{578} (\mn@eprint
  {arXiv} {1807.04288})

\bibitem[\protect\citeauthoryear{Springel}{Springel}{2010}]{Springel2010}
Springel V.,  2010, \mn@doi [MNRAS] {10.1111/j.1365-2966.2009.15715.x}, 401,
  791

\bibitem[\protect\citeauthoryear{Springel, {Di Matteo}  \& Hernquist}{Springel
  et~al.}{2005}]{Springel2005}
Springel V.,  {Di Matteo} T.,   Hernquist L.,  2005, \mn@doi [MNRAS]
  {10.1111/j.1365-2966.2005.09238.x}, 361, 776

\bibitem[\protect\citeauthoryear{Thornton, Gaudlitz, Janka  \&
  Steinmetz}{Thornton et~al.}{1998}]{Thornton1998}
Thornton K.,  Gaudlitz M.,  Janka H.,   Steinmetz M.,  1998, \mn@doi [ApJ]
  {10.1086/305704}, 500, 95

\bibitem[\protect\citeauthoryear{Trebitsch, Volonteri, Dubois  \&
  Madau}{Trebitsch et~al.}{2018}]{Trebitsch2017}
Trebitsch M.,  Volonteri M.,  Dubois Y.,   Madau P.,  2018, \mn@doi [MNRAS]
  {10.1093/mnras/sty1406}

\bibitem[\protect\citeauthoryear{Vasudevan \& Fabian}{Vasudevan \&
  Fabian}{2007}]{Vasudevan2007}
Vasudevan R.~V.,  Fabian A.~C.,  2007, \mn@doi [MNRAS]
  {10.1111/j.1365-2966.2007.12328.x}, 381, 1235

\bibitem[\protect\citeauthoryear{Vogelsberger, Genel, Sijacki, Torrey, Springel
   \& Hernquist}{Vogelsberger et~al.}{2013}]{Vogelsberger2013}
Vogelsberger M.,  Genel S.,  Sijacki D.,  Torrey P.,  Springel V.,   Hernquist
  L.,  2013, \mn@doi [MNRAS] {10.1093/mnras/stt1789}, 3067, 3031

\bibitem[\protect\citeauthoryear{Vogelsberger et~al.,}{Vogelsberger
  et~al.}{2014a}]{Vogelsberger2014}
Vogelsberger M.,  et~al., 2014a, \mn@doi [MNRAS] {10.1093/mnras/stu1536}, 444,
  1518

\bibitem[\protect\citeauthoryear{Vogelsberger, Zavala, Simpson  \&
  Jenkins}{Vogelsberger et~al.}{2014b}]{Vogelsberger2014a}
Vogelsberger M.,  Zavala J.,  Simpson C.,   Jenkins A.,  2014b, \mn@doi [MNRAS]
  {10.1093/mnras/stu1713}, 444, 3684

\bibitem[\protect\citeauthoryear{Wadepuhl \& Springel}{Wadepuhl \&
  Springel}{2011}]{Wadepuhl2010}
Wadepuhl M.,  Springel V.,  2011, \mn@doi [MNRAS]
  {10.1111/j.1365-2966.2010.17576.x}, 410, 1975

\bibitem[\protect\citeauthoryear{Wagner, Umemura  \& Bicknell}{Wagner
  et~al.}{2013}]{Wagner2013}
Wagner A.~Y.,  Umemura M.,   Bicknell G.~V.,  2013, \mn@doi [ApJ]
  {10.1088/2041-8205/763/1/L18}, 763, L18

\bibitem[\protect\citeauthoryear{Whitaker et~al.,}{Whitaker
  et~al.}{2011}]{Whitaker2011}
Whitaker K.~E.,  et~al., 2011, \mn@doi [ApJ] {10.1088/0004-637X/735/2/86}, 735,
  86

\bibitem[\protect\citeauthoryear{York}{York}{2000}]{York2000}
York D.~G.,  2000, \mn@doi [AJ] {10.1086/301513}, 120, 1579

\makeatother
\end{thebibliography}
\appendix

\section{Minimum Mass Parameter} \label{min_mass}

Since we have a constant accretion rate, only the energy injection scheme needs to be converged, which does not necessarily require resolving the Bondi radius. The smallest mass scale of the super-Lagrangian refinement is set via the minimum mass parameter $m_\mathrm{min}$. Before setting up our main simulation suite presented in this paper, we ran simulations equivalent to the `SN+AGNBi100' set-up at lower resolution ($m_\mathrm{cell}=2000\ \mathrm{M_{\odot}}$) to test the convergence behaviour of $m_\mathrm{min}$. The bipolar AGN feedback is particularly reliant on gas flows being accurately resolved due to the more complex injection scheme. 

\begin{figure}
\centering
\includegraphics[width=0.49\columnwidth]{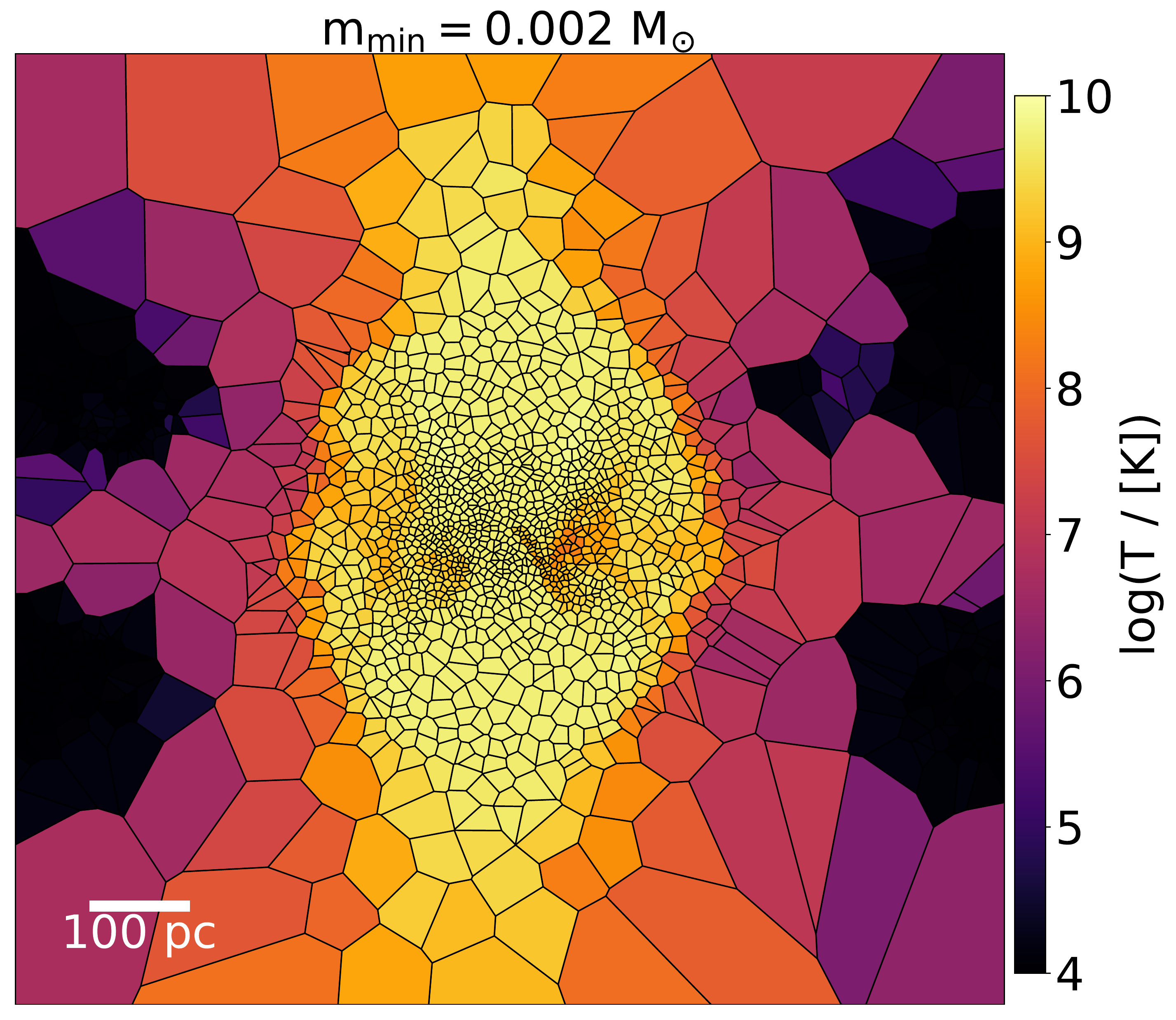}
\includegraphics[width=0.49\columnwidth]{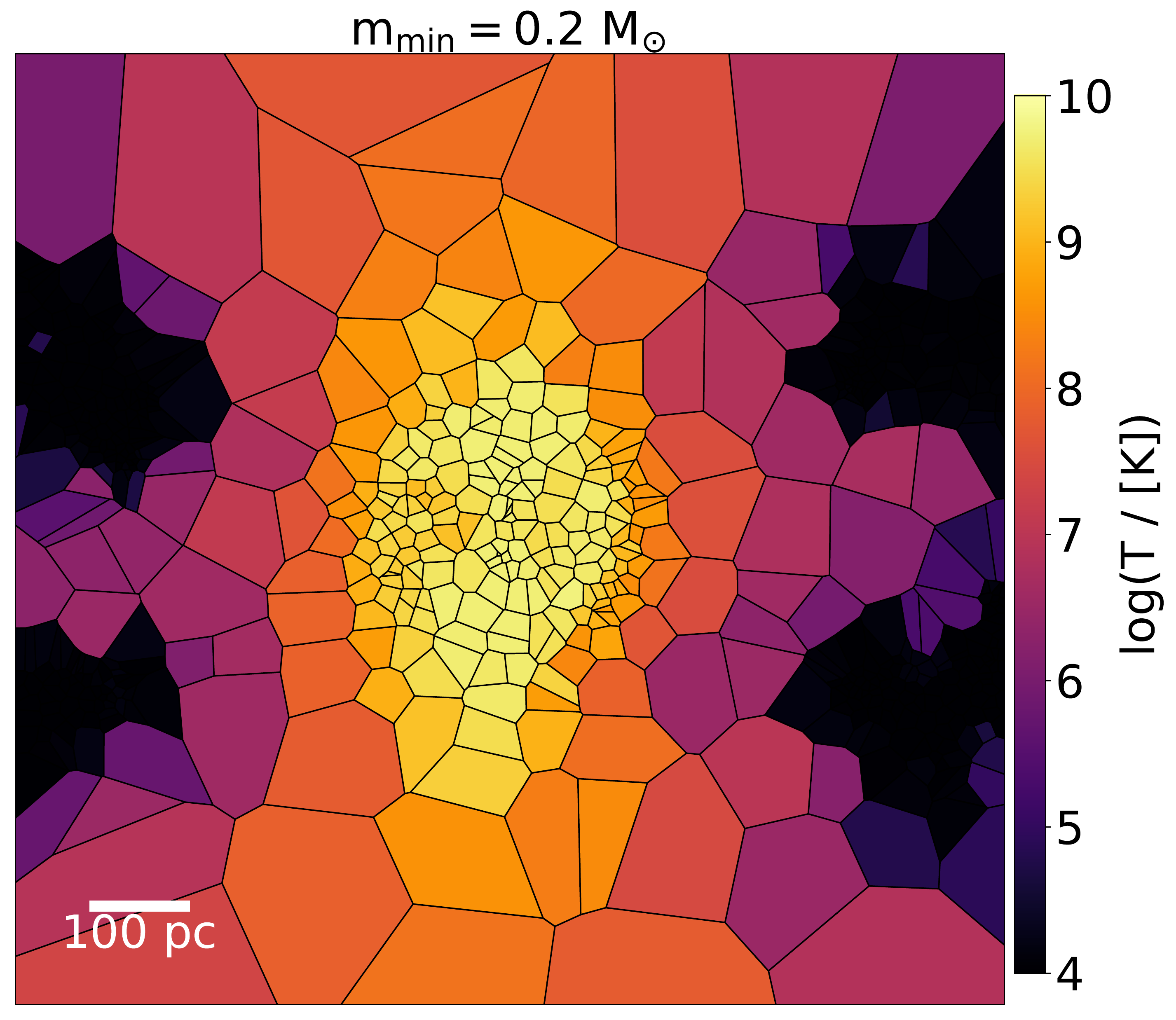}\\
\includegraphics[width=0.49\columnwidth]{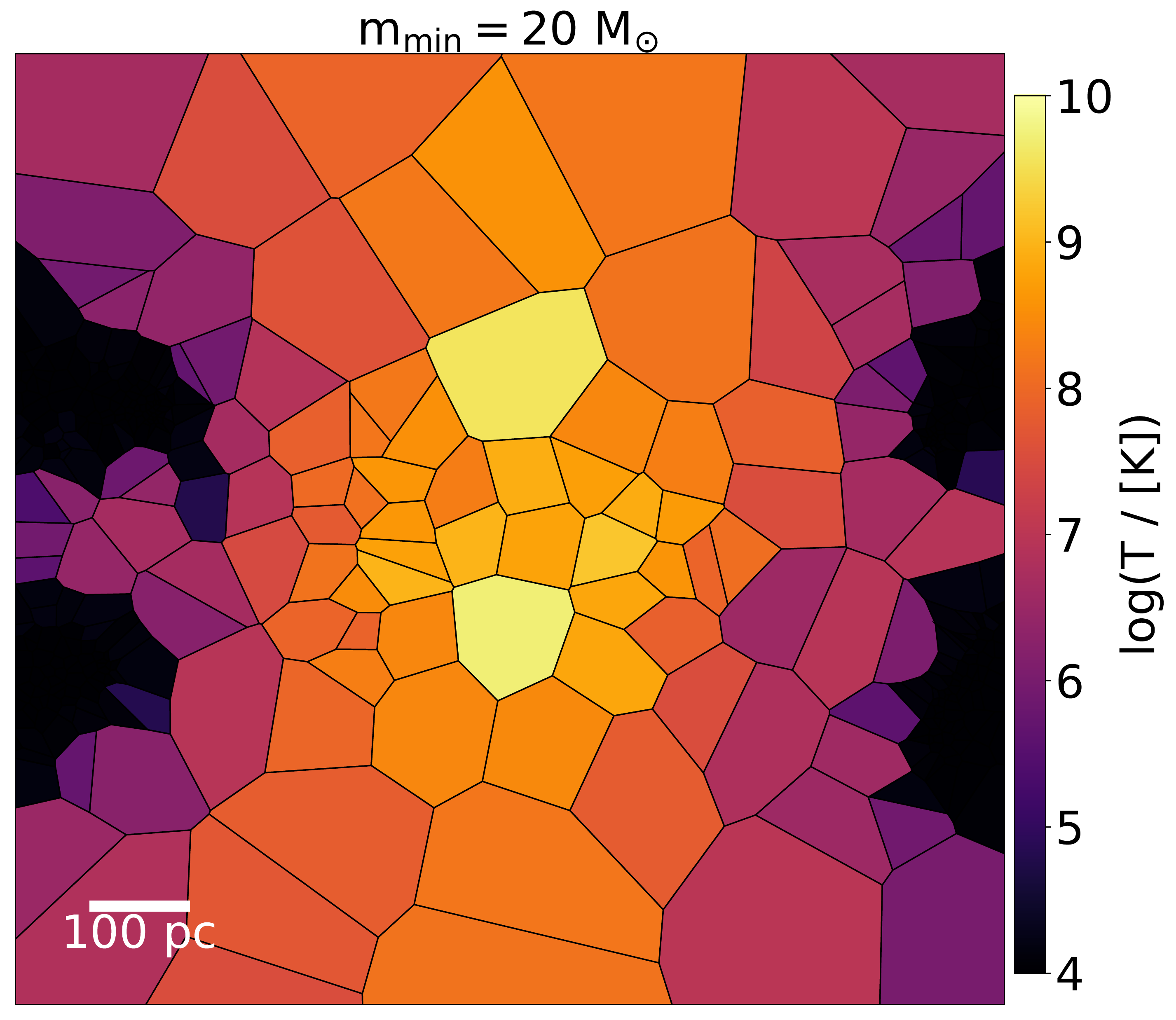}
\includegraphics[width=0.49\columnwidth]{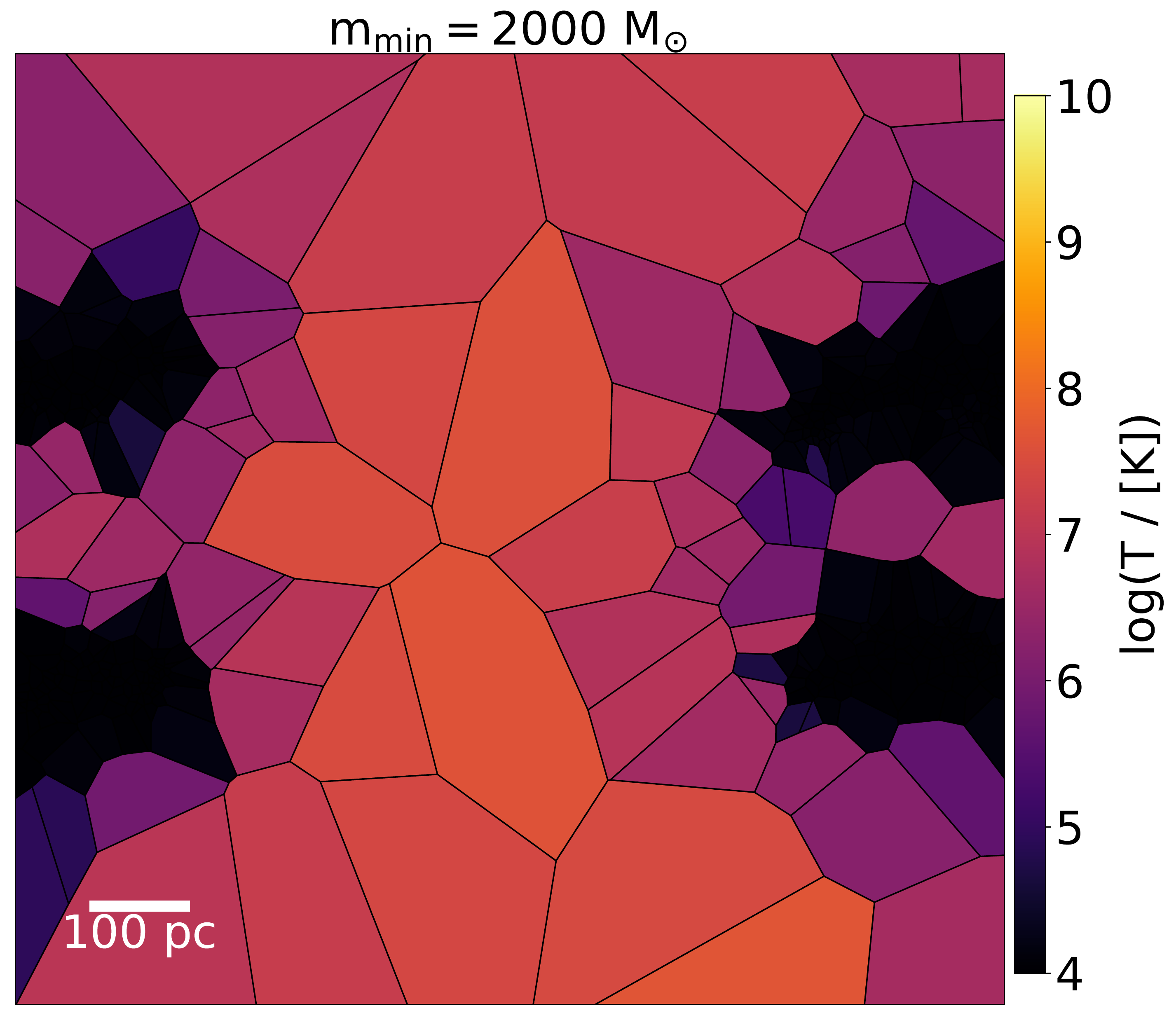}
\caption{$1.0 \times 1.0 \times 0.4$  kpc projections of the Voronoi mesh for the different convergence test runs ($m_\mathrm{min}=0.002,0.2,20,2000\ \mathrm{M_{\odot}}$, where the last one is simply a run without super-Lagrangian refinement)  at $t=25$ Myr. Colour coding indicates the temperature of the gas cells. Without the refinement scheme, the gas overcools. As the minimum mass is decreased, there is a sufficient number of cells within the bipolar cone region and cells retain high thermal energy content. }
\label{fig:Voronoi_convergence}
\end{figure}

Projections of the Voronoi mesh for the four different runs ($m_\mathrm{min}=0.002,0.2,20,2000\ \mathrm{M_{\odot}}$, where the last one is simply a run without super-Lagrangian refinement) at $t=25$ Myr are shown in Figure \ref{fig:Voronoi_convergence}. The colour coding indicates the gas cell temperature. The bipolar cone structure is clearly visible for the lowest minimum mass run, $m_\mathrm{min}=0.002\ \mathrm{M_{\odot}}$, and also discernible for $m_\mathrm{min}=0.2\ \mathrm{M_{\odot}}$. There is no clear cone structure for the other two cases. 

These mesh visualisations illustrate one of the main issues with not accurately resolving energy injection for any type of feedback. Massive gas cells, in general, have a higher absolute energy content. If the amount of energy injected into the cell is small compared to that overall energy content, the temperature change will be small and the gas cell will not get moved to a smaller time-step. If the gas cell's time-step is longer than the cooling time, the thermal energy is lost immediately.  For the bipolar feedback scheme, additional considerations need to be taken into account. In particular with sustained, powerful feedback, the conical energy injection region can get almost completely depleted so that only a few cells actually have their centres in the injection region. In extreme cases, there are no gas cells in the `allowed' bipolar cone region so that the energy cannot get injected. At $t=25$ Myr with $m_\mathrm{min}=20\ \mathrm{M_{\odot}}$, two massive cells located in the cone region have received the majority of the energy, whilst the cells in the very centre of the galaxy have received no energy as they are not within the cone region. The $m_\mathrm{min}=2000\ \mathrm{M_{\odot}}$ overcools with no cells above $10^8$ K in the inner region. This shows that to heat the ISM and CGM from the galaxy centre outwards with a conical geometry we require a minimum mass of at most $0.2\ \mathrm{M_{\odot}}$.

\begin{figure}
\centering
\includegraphics[width=\columnwidth]{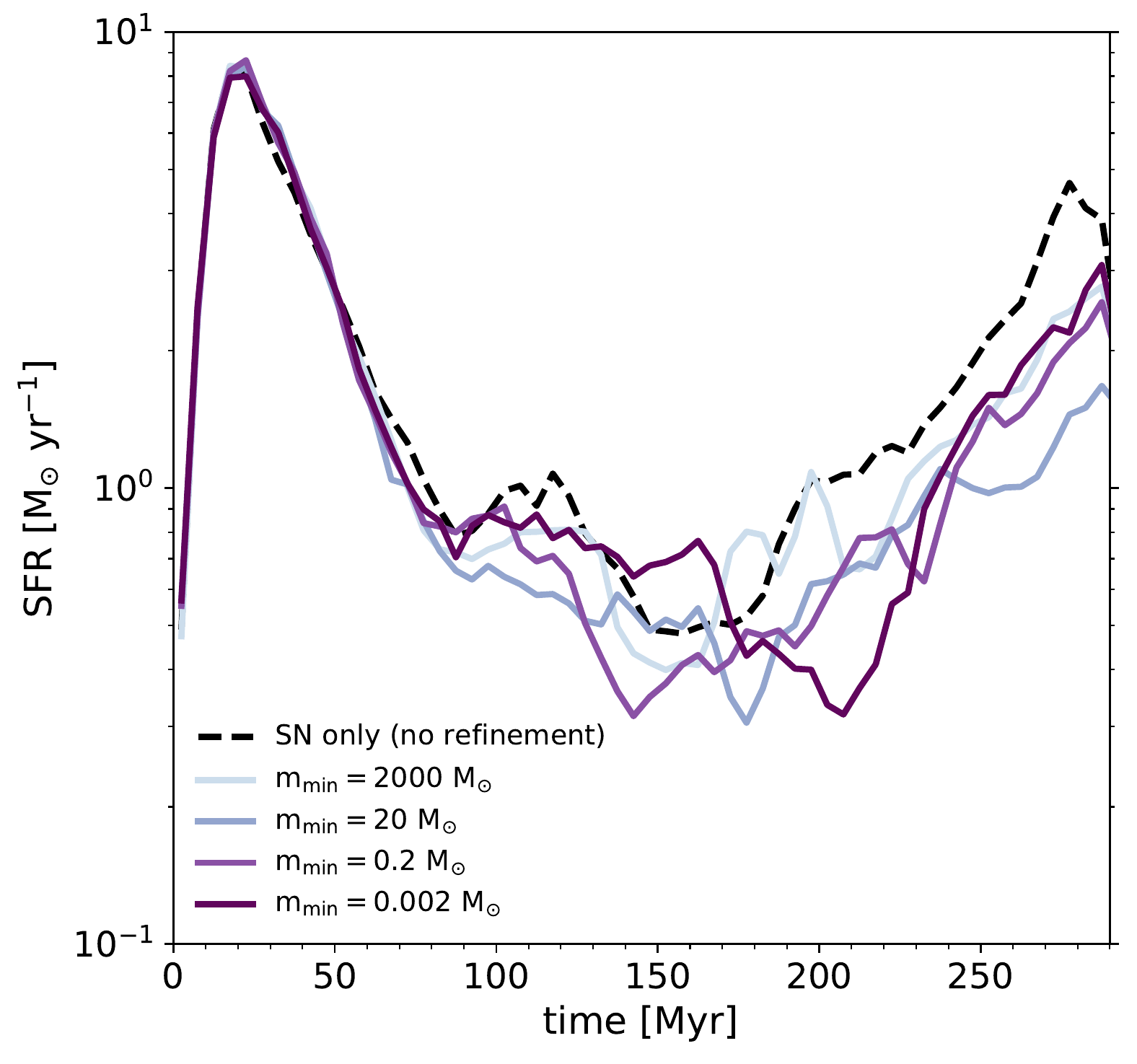}
\caption{SFRs against time for the low resolution AGN convergence test runs. For comparison the SN only run is also shown. The SFRs of all AGN runs are in good agreement, within the expected fluctuations. From $t=200$ Myr onwards, the SFRs of the AGN runs are systematically lower than the SFRs of the SN run.}
\label{fig:sfr_convergence}
\end{figure}

\begin{figure*}
\centering
\includegraphics[width=\textwidth]{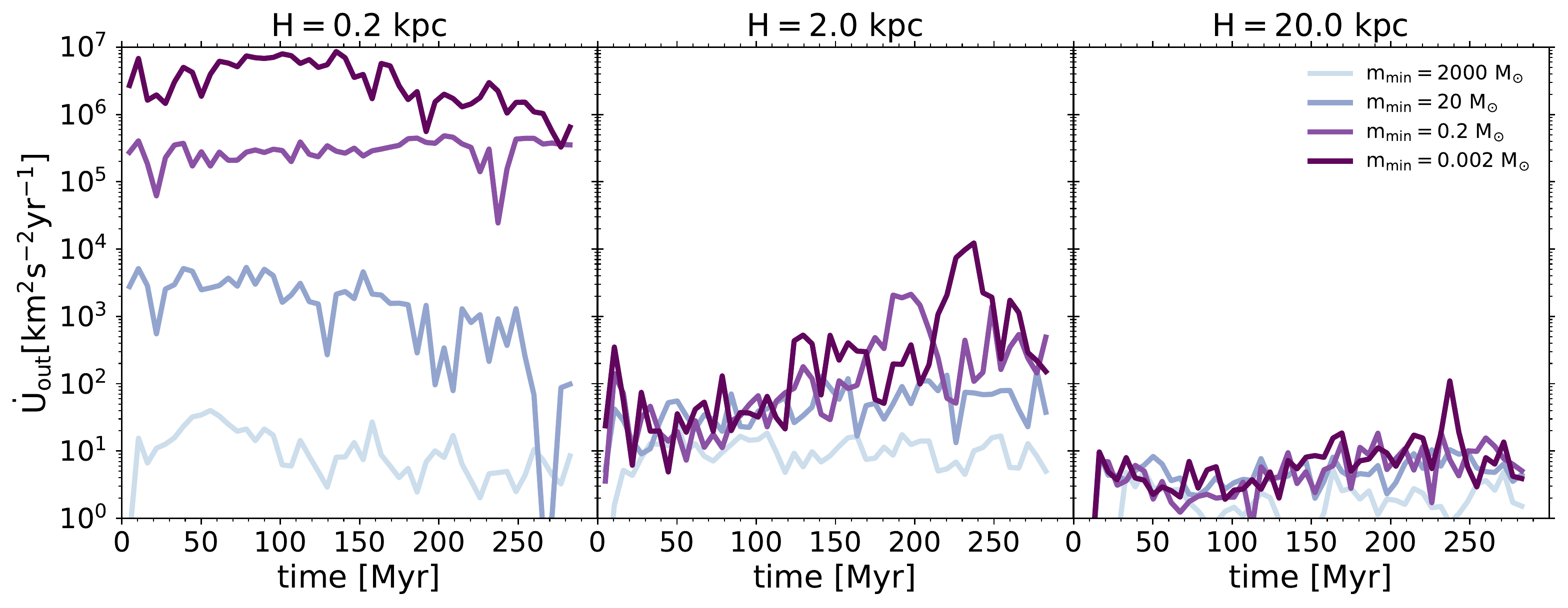}
\caption{Thermal energy outflow rates against time for the low resolution AGN convergence test runs calculated at target heights $\mathrm{H=0.2\ kpc}$  (left panel),  $\mathrm{H=2.0\ kpc}$ (middle panel), and $\mathrm{H=20.0\ kpc}$ (right panel). Thermal energy outflow rates are significantly higher with super-Lagrangian refinement and decreasing minimum mass. This effect is particularly significant at small scales.}
\label{fig:outflows_convergence}
\end{figure*}

Next, we gauge the effect of the minimum mass parameter more quantitatively, by assessing star formation and outflow convergence. Figure \ref{fig:sfr_convergence} shows the SFRs for the four bipolar feedback convergence runs. For comparison the SFR of the equivalent SN only run is also shown. Overall, the SFRs for the bipolar runs are in good agreement - within the expected fluctuations. From $t=200$ Myr onwards, all of the bipolar runs are systematically offset from the SN only run though this offset is small, similarly to the equivalent high-resolution runs (cf. Figure \ref{fig:sfr}).

Figure \ref{fig:outflows_convergence} shows the thermal energy outflow rate $\dot{U}_\mathrm{out}$ at three target heights, $H=0.2$ kpc, $H=2.0$ kpc, and $H=20.0$ kpc, for a slice of thickness $\mathrm{d}z=200$ pc. There are two significant differences to note. Firstly, the driving of outflows on large scales is slightly delayed without extra refinement, suggesting that we need the super-Lagrangian refinement to efficiently drive outflows. Secondly, there are significant differences in the thermal energy outflow rates, especially at small scales. The lower the minimum mass, the more efficiently thermal energy is transported away from the black hole. Even at large scales, $H=20.0$ kpc, there is still a factor five difference when the super-Lagrangian refinement scheme is employed. We also inspected velocity histograms split by temperature and found that there is more fast and very hot gas with additional refinement. For the overall mass outflow rate and average outflow velocity, there are no significant differences between the schemes, as these are dominated by the slower moving warm phase, present in all of the runs.

Overall, we conclude that for the thermal outflow properties to be converged, we need a minimum mass in the refinement region of at most $0.2\ \mathrm{M_{\odot}}$, even though for getting SFRs converged this is not crucial, as the cold star-forming phase is not affected.
\vspace{-4ex}
\section{Super-Lagrangian Refinement Scheme and Feedback} \label{SuperLagRef}

The super-Lagrangian refinement scheme sets a range of allowed cell radii for the gas cells within the refinement radius $R_\mathrm{ref}$. Recall that the refinement radius is set equal to the black hole smoothing length so that we will have a larger refinement region for lower central densities (see Section \ref{AGNmodel}). The maximum cell radius at the boundary of the refinement region is set proportional to the refinement region radius. Both radii are related to the target gas cell mass via the boundary cell density and the average refinement region density, respectively. However, these two radii will only be proportional to each other, if we assume the ratio of these two densities to be constant. For example, if we assume these two densities to be equal, the ratio should be approximately 0.3. 

\begin{figure*}
\centering
\includegraphics[width=\textwidth]{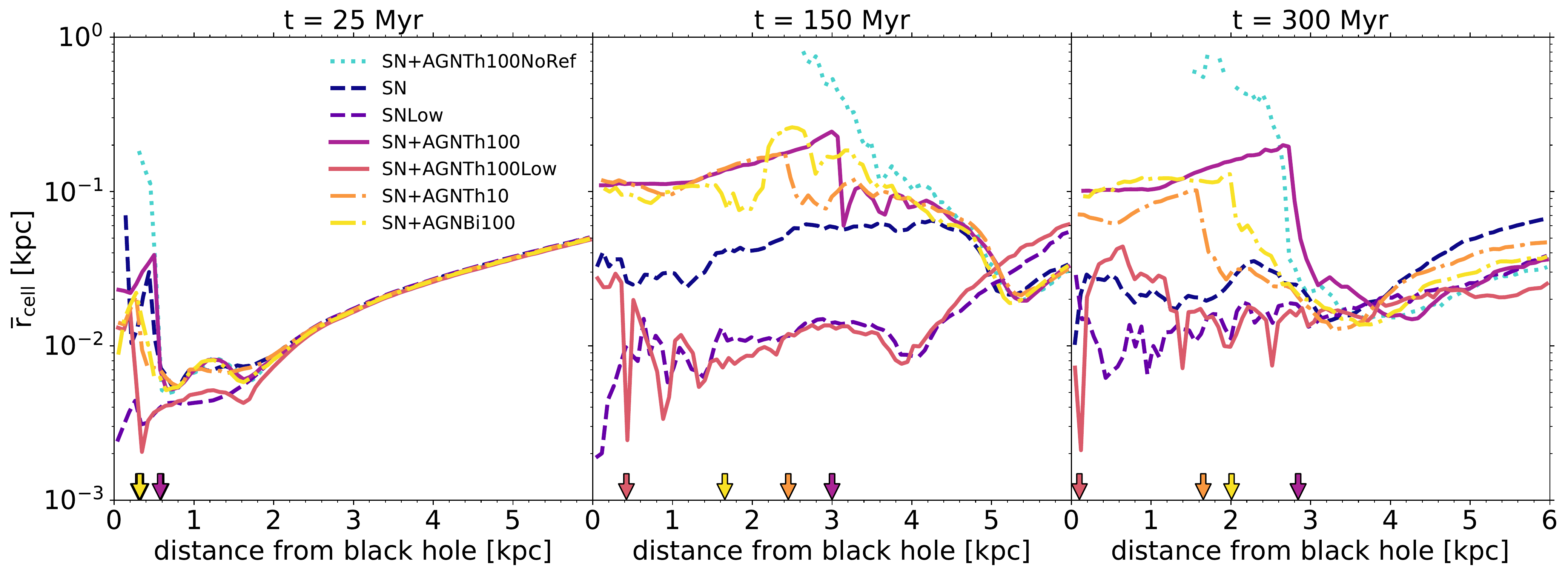}
\caption{Mean gas cell radii as a function of distance from the black hole for all feedback runs at $t=25, 150, 300$ Myr. For comparison, we also show the cell sizes for a run with the same set-up as `SN+AGNTh100' but without super-Lagrangian refinement around the black hole. The respective sizes of the refinement region are indicated by the coloured arrows. The AGN energy injection significantly lowers the density of the central region. Without additional refinement, central cell sizes reach $\sim 1$ kpc and we therefore do not have any gas cell centres near the black hole at late times. The super-Lagrangian refinement scheme counteracts this cell size inflation and the region around the black hole stays well-resolved.}
\label{fig:cell_sizes}
\end{figure*}

We expect the average refinement region density to be lower than the value at the boundary, since the latter is less affected by feedback, and therefore choose a fiducial value of $r_\mathrm{cell,boundary}=0.1R_\mathrm{ref}$. With strong feedback, this value might need to be even lower, and with weak feedback, leaving the refinement region with a relatively uniform density, this value would need to be higher. We do not want to overtune our simulations, so we use $r_\mathrm{cell,boundary}=0.1R_\mathrm{ref}$ for all of our runs. Figure \ref{fig:cell_sizes} shows the cell size distribution for the low and high star formation efficiency runs at $t=25,150,300$ Myr. We plot average cell size against distance from the black hole to check whether the gas cells within the smoothing length have been properly refined. For comparison, we also show the cell sizes for a run with the same set-up as `SN+AGNTh100' but without super-Lagrangian refinement around the black hole.

Firstly, we note that the cells in the refinement region, whilst much smaller in mass, are not necessarily smaller in volume. This is because in our simulations, the refinement scheme mainly counteracts the strong AGN feedback inflating the cells around the black hole, leading to relatively smooth cell sizes, rather than ultra refined ones. For both of the high-luminosity thermal runs, there are sharp transitions in cell size at the refinement region border. For `SN+AGNTh100', these always go to larger cell sizes as the central region has very low densities due to the early, high SN activity and aggressive AGN feedback. For the `SN+AGNTh100Low' run, there is a transition to smaller cell sizes. The transitions for the `SN+AGNBi100' and `SN+AGNTh10' run are more regularized, since the former does not inject energy directly into the disc, and the latter is at lower luminosity and less aggressive. However, note that without super-Lagrangian refinement, the cell size transition is even more extreme, sharply rising to $\sim 1\ \mathrm{kpc}$. Furthermore, due to the inflated cell sizes, there are no gas cell centres in the central kpc at late times.

Also note that there are similar spatial resolution issues with any kind of extreme feedback, e.g. the strong SN feedback in the `SN' run also leads to larger cell sizes in the centre, in particular at early times. In addition, there are some small discontinuities simply from using up the central dense gas in star formation which introduces a bias towards larger cell sizes at the centre of the galaxy.

\begin{figure*}
\centering
\includegraphics[width=0.245\textwidth]{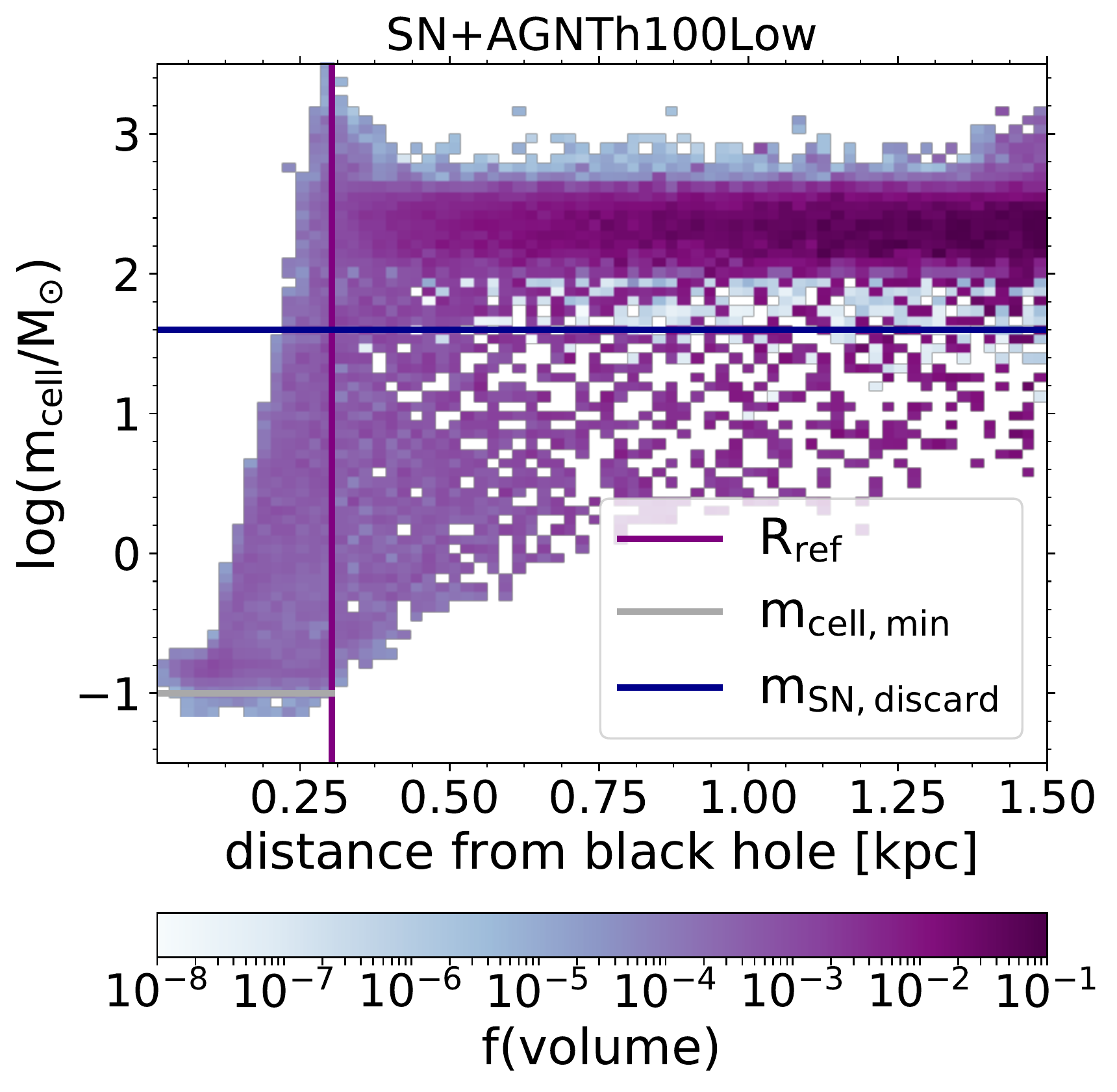}
\includegraphics[width=0.245\textwidth]{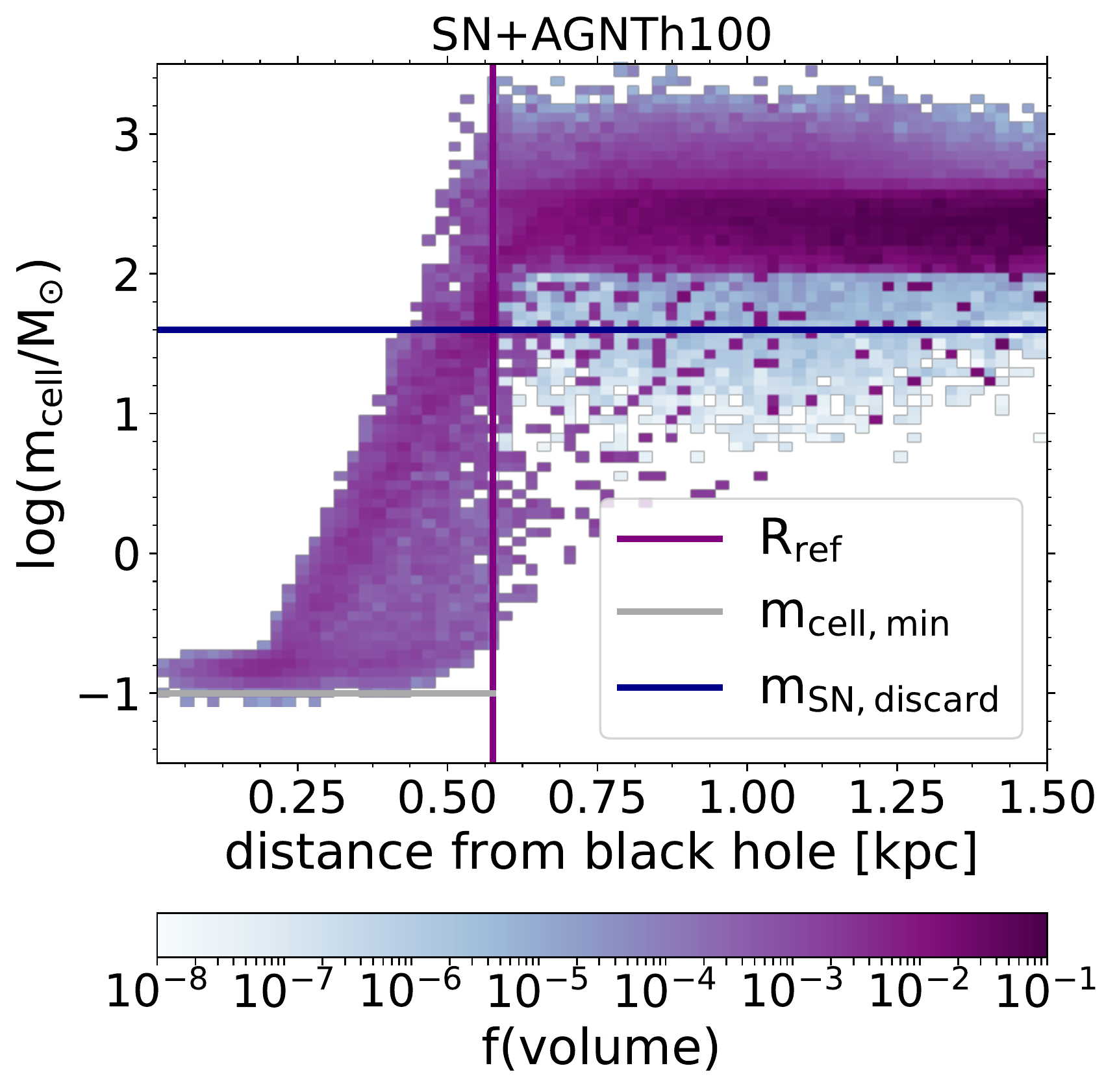}
\includegraphics[width=0.245\textwidth]{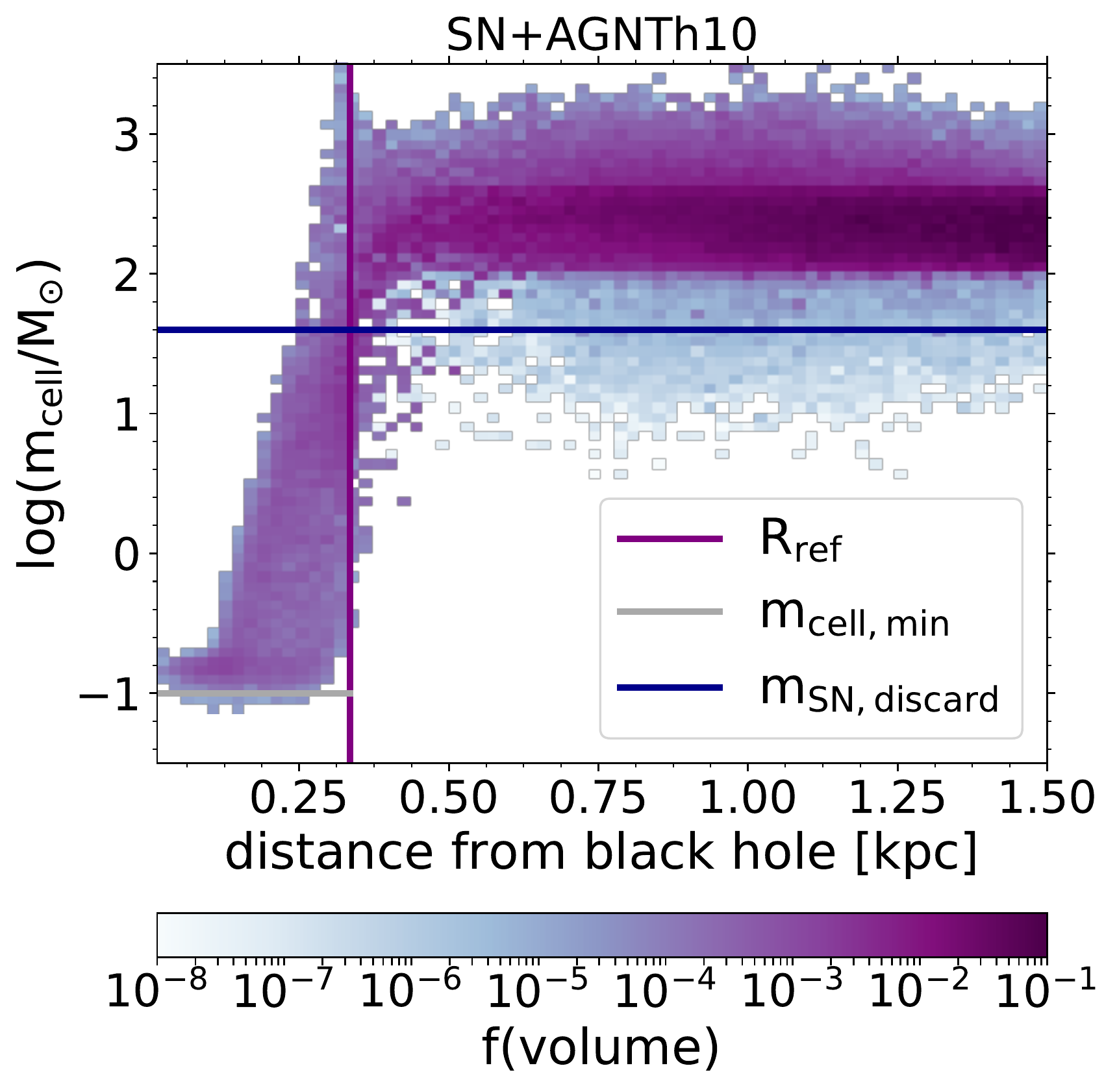}
\includegraphics[width=0.245\textwidth]{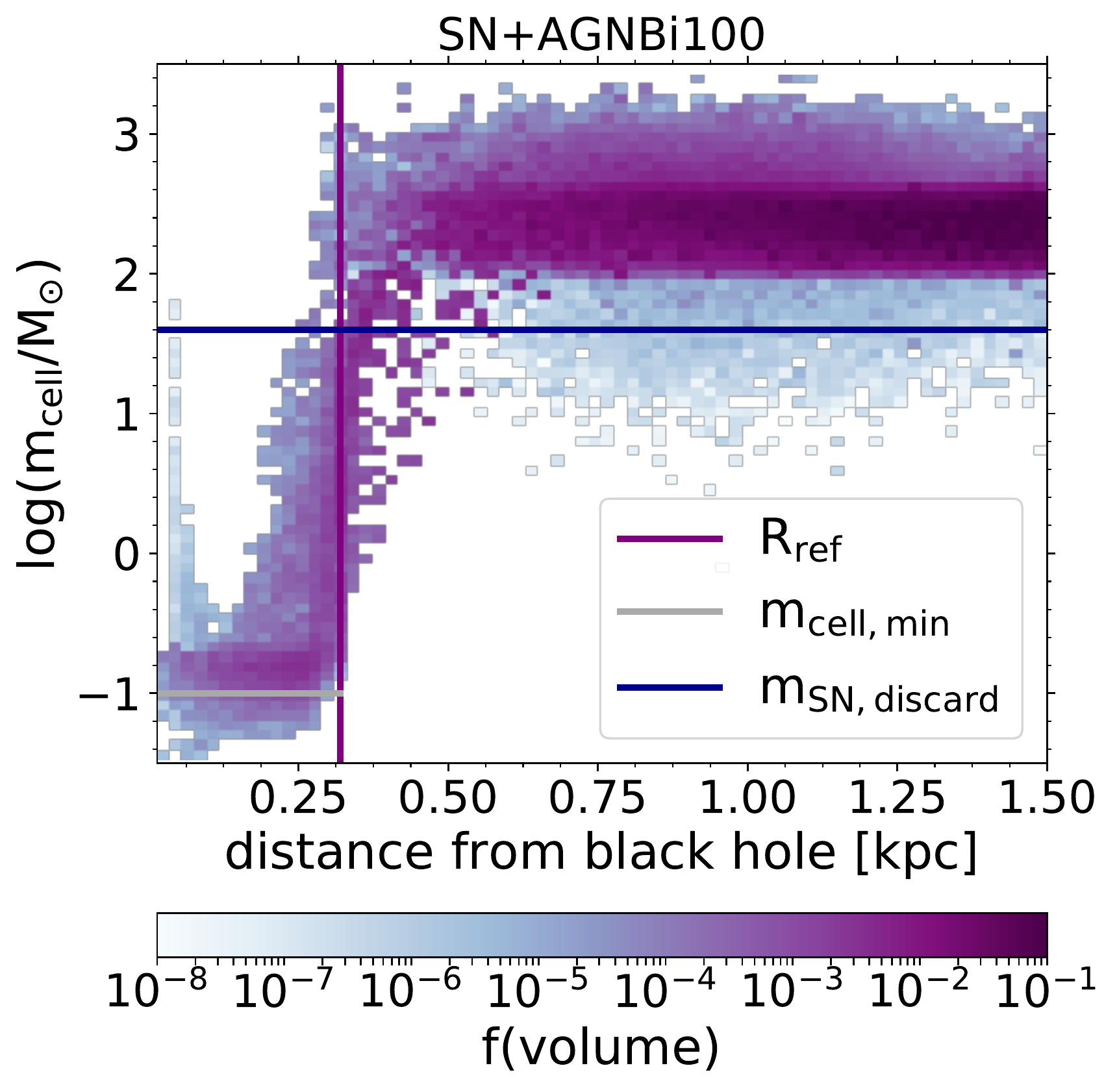}
\caption{Volume-weighted distribution of gas cell masses as a function of distance from the black hole for the runs with AGN feedback runs at $t=25$ Myr. The border of the refinement region is indicated in purple, the minimum gas mass in grey, and the minimum mass for hosting a SN in blue.}
\label{fig:cell_masses}
\end{figure*}

\begin{figure*}
\centering.
\includegraphics[width=\textwidth]{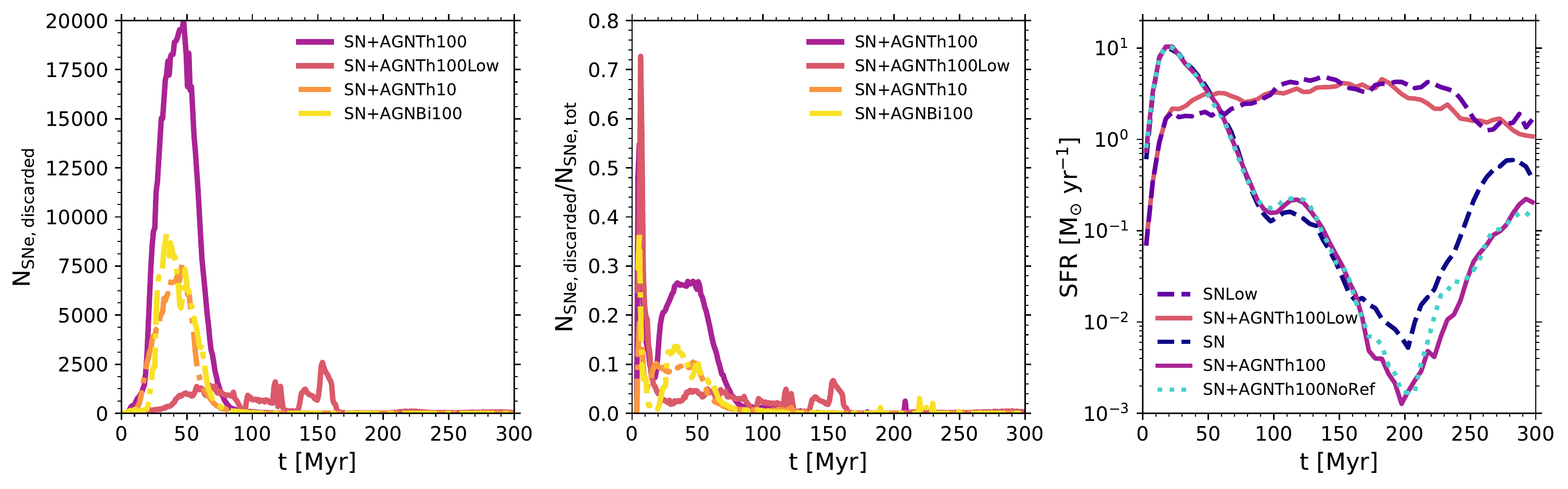}
\caption{ The impact of discarding SNe in super-refined cells ($m_\mathrm{cell}<0.2m_\mathrm{target}$). \textit{Left panel}: number of discarded SNe vs. time. \textit{Middle panel}: fraction of discarded SNe against time for the AGN feedback runs. \textit{Right panel}: SFRs for the most aggressive AGN feedback models (thermal feedback at the Eddington luminosity) and the SN only runs as well as the thermal AGN feedback run without additional refinement (`SN+AGNTh100NoRef'), for the latter two virtually no SNe are discarded as we do not use the super-Lagrangian refinement scheme. Note that even though a significant number of SNe get discarded at early times (between 6 and 28 per cent), the SFRs are not affected. This is because the inner region, where we discard the SNe, is already heated up by the AGN activity so additional SNe would not make much of a difference here.  }
\label{fig:sn_discarded}
\end{figure*}

\citet{Curtis2015} already noticed this issue with constant accretion schemes and find that if they go higher than ten per cent of the initial Eddington rate, the gas gets blown away on a short time-scale. As we only run our simulations for up to 300 Myr, we still have enough gas to feed the black hole, however this is an important caveat of the constant accretion scheme that needs to be kept in mind when testing extreme cases. Though this is beyond the scope of this paper exploring the effects of additional regularisation mechanisms (e.g. setting a maximum volume ratio for neighbouring cells) would be desirable for a future study. 

Figure \ref{fig:cell_masses} shows the volume-weighted distribution of cell masses in the central region at $t=25\ \mathrm{Myr}$.  The minimum allowed gas cell mass $10^{-1}\ \mathrm{M_{\odot}}$ (cells have to be within a factor two of the minimum mass $2\times10^{-1}\ \mathrm{M_{\odot}}$) is indicated as a horizontal grey line. A significant fraction of gas cells in the refinement region reach the minimum mass, which we will have to keep in mind for the interpretation of our results.  For the bipolar injection scheme we also have some cells close to the black hole with much higher masses, these are the cells outside the cone region which are unaffected by feedback and therefore have a much higher density. This figure also shows how many gas cells fall below the cut-off mass for hosting SNe $m_\mathrm{SN,discard}$ (indicated as a blue horizontal line). Outside of the refinement region only a negligible amount of gas cells is below the SN threshold mass suggesting that SN rates should not be severely affected by this criterion.

To make this claim more quantitative, we show the number of discarded SNe as well as the ratio between discarded and total number of SNe for all AGN runs in Figure \ref{fig:sn_discarded}. For comparison, we also show the SFRs of the high-luminosity thermal AGN runs, including the high $\epsilon_\mathrm{SF}$ run without additional refinement, and the SN only runs. Most of the SNe are discarded at early times, however, as can be seen from the SFRs, this does not affect global star formation. For the first $\sim$ 25 Myr, there is a rapid gas collapse from the initial conditions and SNe are not effective. Therefore losing SNe at early times will have a small effect. Overall, the discarded fraction is approximately 2 per cent for `SN+AGNTh100Low', 6 per cent for `SN+AGNTh10', 7 per cent for `SN+AGNBi100', and 18 per cent for `SN+AGNTh100'. For the first three, the discarded fraction is not significant but nevertheless non-negligible and could have a small effect on total feedback energetics. The high discarded fraction for the `SN+AGNTh100' run is largely due to the aggressive constant isotropic feedback, heating up and dispersing the gas around the black hole. However, this aggressive feedback also means that the inner region, where we discard the SNe, is significantly heated up anyway so additional SNe would not make much of a difference here, as can be seen from the negligible difference in SFRs between `SN+AGNTh100' and `SN+AGNTh100NoRef'. Also note that the same SN event might get discarded repeatedly so our discarded SNe fractions represent an upper limit.
\end{document}